%% file: main.tex
\documentclass[letterpaper,11pt]{article}

\usepackage{arxiv}
\usepackage{amsmath}
\usepackage{amssymb}
\usepackage[numbers,sort&compress]{natbib}
\usepackage[font=small]{caption}
\usepackage{graphicx}
\usepackage{xcolor}
\usepackage[hidelinks]{hyperref}
\usepackage[caption=false]{subfig}
\usepackage{booktabs}
\usepackage{tabularx}
\usepackage{placeins}
\usepackage[capitalise,nameinlink,noabbrev]{cleveref}

\numberwithin{equation}{section}
\setkeys{Gin}{trim=4 4 4 4,clip}
\input{macros}

\renewcommand{\headeright}{Preprint}
\renewcommand{\undertitle}{Preprint}
\renewcommand{\shorttitle}{Flood Monitoring Objectives}
\title{Matching Urban Flood Sensor Placement to Monitoring Objectives Using Bayesian Optimal Experimental Design}
\hypersetup{
  pdftitle={Matching Urban Flood Sensor Placement to Monitoring Objectives Using Bayesian Experimental Design},
  pdfauthor={Chen Cheng, Vinh Ngoc Tran, Jiayuan Dong, Sarah Whitaker, Shannon Bergt, John Ziker, Valeriy Y. Ivanov, and Xun Huan},
  pdfsubject={arXiv preprint},
  pdfkeywords={Bayesian optimal experimental design, flood inundation, goal-oriented design, monitoring networks, sensor placement, uncertainty quantification}
}

\date{}
\author{
  \textbf{Chen Cheng$^{1}$,
  Vinh Ngoc Tran$^{2,3}$,
  Jiayuan Dong$^{1}$,
  Sarah Whitaker$^{4}$}\\[0.35em]
  \textbf{Shannon Bergt$^{5}$,
  John Ziker$^{4}$,
  Valeriy Y. Ivanov$^{2}$,
  Xun Huan$^{1}$}\thanks{Correspondence: Xun Huan (\href{mailto:xhuan@umich.edu}{xhuan@umich.edu}, \href{https://uq.engin.umich.edu}{https://uq.engin.umich.edu})}\\[0.75em]
  \normalfont
  $^{1}$Department of Mechanical Engineering, University of Michigan, Ann Arbor, MI 48109, United States\\
  \normalfont $^{2}$Department of Civil and Environmental Engineering, University of Michigan,\\
  Ann Arbor, MI 48109, United States\\
  \normalfont $^{3}$Computational Sciences and Engineering Division, Oak Ridge National Laboratory,\\
  Oak Ridge, TN 37830, United States\\
  \normalfont $^{4}$Department of Anthropology, Boise State University, Boise, ID 83725, United States\\
  \normalfont $^{5}$Department of Defense, U.S. Army Garrison Detroit Arsenal, Warren, MI 48091, United States
}

\begin{document}
\maketitle

\begin{abstract}
Flood-monitoring sensors are often placed according to coverage, access, or expected inundation. However, the value of a measurement depends on the prediction or decision it is intended to inform. Using tRIBS-Urban simulations and a neural-network surrogate of the August 2014 metropolitan Detroit flood, we examine how this learning target changes single-sensor placement. Across 2,576 candidate locations, we compare parameter-oriented optimal experimental design (PO-OED), which values expected information gain (EIG) about model parameters, with goal-oriented optimal experimental design (GO-OED), which values EIG about specified flood predictions. We also examine how parameter EIG evolves during the event, and illustrate that parameter learning translates unevenly into reductions in predictive uncertainty across locations and lead times. Under GO-OED, point-depth targets favor nearby locations, whereas regional-average and regional maximum-depth targets can favor nonlocal locations. Weighted multi-point objectives retain similar broad spatial patterns, although their computed max-EIG locations differ. Public geospatial data further provide illustrative feasibility and contextual classifications for deployment screening. These results show how monitoring objectives shape sensor placement in optimal experimental design, and motivates an objective-first workflow that defines the intended prediction and priorities, applies field-verified restrictions, and ranks locations by EIG.
\end{abstract}


\input{sections/01_introduction}

\input{sections/02_problem}
\input{sections/03_procedure}
\input{sections/04_results}
\input{sections/05_discussion}
\input{sections/06_conclusion}

\section*{Acknowledgments}
This work relates to the Department of the Navy Award No.\ N000142512411 issued by the Office
of Naval Research.

\section*{Conflict of Interest}
The authors declare no conflicts of interest.

\section*{Data Availability Statement}
The data that support the findings of this study are available from the corresponding author upon reasonable request.

\bibliographystyle{abbrvurl}
\bibliography{references}

\clearpage
\input{sections/07_appendix}

\end{document}

%% file: macros.tex
\newcommand{\design}{\xi}
\newcommand{\designset}{\mathcal{D}}

\newcommand{\Param}{\Theta}
\newcommand{\param}{\theta}

\newcommand{\EE}{\mathbb{E}}

\newcommand{\argmax}{\operatornamewithlimits{argmax}}
\newcommand{\DKL}{D_{\mathrm{KL}}}

%% file: sections/01_introduction.tex
\section{Introduction}

Flood-risk management uses inundation models to predict water depth, flow pathways, and flood timing~\citep{Teng2017}. These predictions carry uncertainty because precipitation forcing, topography, initial conditions, model parameters, and model structure are not perfectly known~\citep{Beven2016,Savage2016,Pappenberger2008,Grimaldi2019}. 
The uncertainties propagate through nonlinear flood dynamics and their interactions become especially complex in urban environments, where buildings obstruct overland flow, roadways convey runoff, and drainage systems transmit flood effects beyond areas receiving local rainfall~\citep{Ivanov2021,Tran2024,Balaian2024}. 
Monitoring can reduce uncertainty, but the value of a prospective measurement depends on the prediction or decision it is intended to inform. Flood-monitoring design must therefore define this target before deciding what, where, and when to measure, because a design is optimal only relative to the monitoring objective it encodes.

Existing flood-monitoring studies explore several different notions of value. Data-assimilation results show that sensor location, network configuration, and observation timing affect forecast performance~\citep{Mazzoleni2018,VanWesemael2019,Dasgupta2021}; other formulations emphasize graph structure, network observability, or multiple operational objectives~\citep{Ogie2017,Farahmand2022,Tien2023}. 
Adaptive sampling has used expected reductions in water-level forecast-error variance at a specified location and time~\citep{Neal2012}. Information-theoretic network designs combine value-of-information and transinformation-entropy scores for node information and redundancy~\citep{Zheng2025}, or maximize mutual information between observations and the distributed sewer-network state under a linear-Gaussian model~\citep{Crowley2025}. Decision-theoretic analyses evaluate expected decision utility~\citep{Alfonso2012} or identify flood-risk uncertainties worth reducing~\citep{Velandia2024}. 
A related study ranks sensor locations using principal-component scores, and uses the resulting measurements for inundation-field reconstruction to support a vision-transformer forecast model~\citep{Tran2026}.
These formulations answer different monitoring questions by emphasizing measurements according to forecast performance, field reconstruction, network coverage or state, node informativeness and redundancy, operational objectives, or expected decision performance.

Bayesian optimal experimental design (OED) makes the learning target explicit by assigning each candidate an expected utility over prospective observations~\citep{Chaloner1995,Ryan2016,Alexanderian2021,Rainforth2024,Strutz2024,Huan2024}. When the utility is the Kullback--Leibler divergence from the posterior to prior, the resulting expected information gain (EIG) measures how strongly a prospective observation is expected to change the distribution of the learning target; equivalently, it is the mutual information between the observation and that target~\citep{Lindley1956}. Covariance-based A- and D-optimality provide simpler alternatives to this full-distribution criterion, with A-optimality minimizing the posterior covariance trace and D-optimality its log determinant. Both characterize posterior contraction through second moments and therefore do not capture higher-order or non-Gaussian structure. Under linear-Gaussian settings, the EIG collapses to D-optimality.

Recent monitoring applications span these choices. An urban-drainage study estimates parameter EIG for sensor placement and model calibration~\citep{Huang2025}, while a tsunami study combines a linear surrogate with Bayesian approximation error and optimizes the A-optimal criterion~\citep{Koval2025}. A tsunami preprint also applies D-optimality to a linear-Gaussian inverse problem of the uncertain source field~\citep{Venkat2026}. These applications, however, all target learning of uncertain parameters or source fields rather than downstream predictions.

Goal-oriented OED (GO-OED) instead makes a predictive quantity of interest (QoI) the learning target. For linear-Gaussian inverse problems, existing approaches include goal-oriented A- and D-optimal criteria based on the posterior covariance of linear QoIs~\citep{Attia2018,Wu2023}.
For nonlinear models, Monte Carlo and density-estimation methods~\citep{Zhong2026} and variational lower bounds~\citep{Shen2025} have been developed to estimate the prediction EIG.
In flood-monitoring, example QoIs include point depths, joint and weighted multiple-point quantities, regional averages, and vectors of regional spatial maxima across several lead times. 
Thus, while conventional parameter-oriented OED (PO-OED) values information about the full parameter vector, GO-OED concentrates on parameter directions that influence the specified QoI. The appropriate criterion therefore depends on whether monitoring is intended to support parameter learning or a particular prediction.
To our knowledge, the flood-monitoring literature has not yet conducted controlled comparisons between PO-OED and GO-OED.

Furthermore, an informative sensor location is not automatically deployable.
Access, safety, power, communications, permission, maintenance, and organizational responsibility can restrict sensor placement~\citep{Adesina2024}. Verified hard constraints should remove inadmissible locations from the design set; graded feasibility and contextual evidence can guide review of the informative candidates that remain. Public geospatial data can support this screening, which would then be followed with site inspection and stakeholder review.

This paper examines how the learning target changes single-sensor placement in a controlled urban flood case study. The case is historically grounded in the August 2014 storm in the I-696--Mound Road corridor of metropolitan Detroit and adopts the tRIBS-Urban domain described in~\citep{Tran2024,Tran2026}. The present analysis uses a neural-network surrogate trained over a six-parameter ensemble.
Across the GO-OED comparisons, the surrogate, parameter prior, precipitation forcing, observation model, observation time, and 2,576 candidate locations remain fixed while only the QoI changes.
The PO-OED analysis also examines several observation times. 
The resulting rankings show that point-depth targets favor nearby locations, regional-average and regional maximum-depth targets can favor nonlocal locations, and multi-point priority scenarios can preserve similar broad spatial patterns while producing different computed max-EIG locations. Synthetic-observation diagnostics further illustrate how parameter uncertainty reduction can transfer unevenly to predictions across locations and lead times, while public geospatial classifications illustrate how feasibility and contextual evidence can organize subsequent deployment screening.

The paper contributes a controlled application and interpretation of established OED methods for flood monitoring, together with a workflow that connects information value to deployment screening. Specifically:
\begin{itemize}
    \item we compare parameter EIG and prediction EIG for single-sensor placement under a common urban flood model and uncertainty specification;

    \item we evaluate point, multi-point, regional-average, and regional maximum-depth targets, showing how target definition and assigned priorities affect high-ranking sensor locations; and

    \item we provide guidance for defining prediction targets, interpreting local and nonlocal rankings, and incorporating hard constraints, relative feasibility, and contextual evidence into deployment review.
\end{itemize}

The remainder of the paper presents the case study and Bayesian design problem in \cref{sec:problem}, the design criteria and computational approximations in \cref{sec:procedure}, the results in \cref{sec:results},  interpretation and monitoring guidance in \cref{sec:discussion}, and conclusions in \cref{sec:conclusion}.

%% file: sections/02_problem.tex
\section{Urban flood case study and monitoring-design problem}\label{sec:problem}

The case study asks where to place one water-depth sensor before an observation is available. This design problem connects four objects: an inundation model with uncertain inputs, the measurement to be collected, the parameter or prediction to be learned, and the candidate sensor location.

\subsection{Study domain and inundation model}

\begin{figure}[!htb]
    \centering
    \includegraphics[width=.92\textwidth]{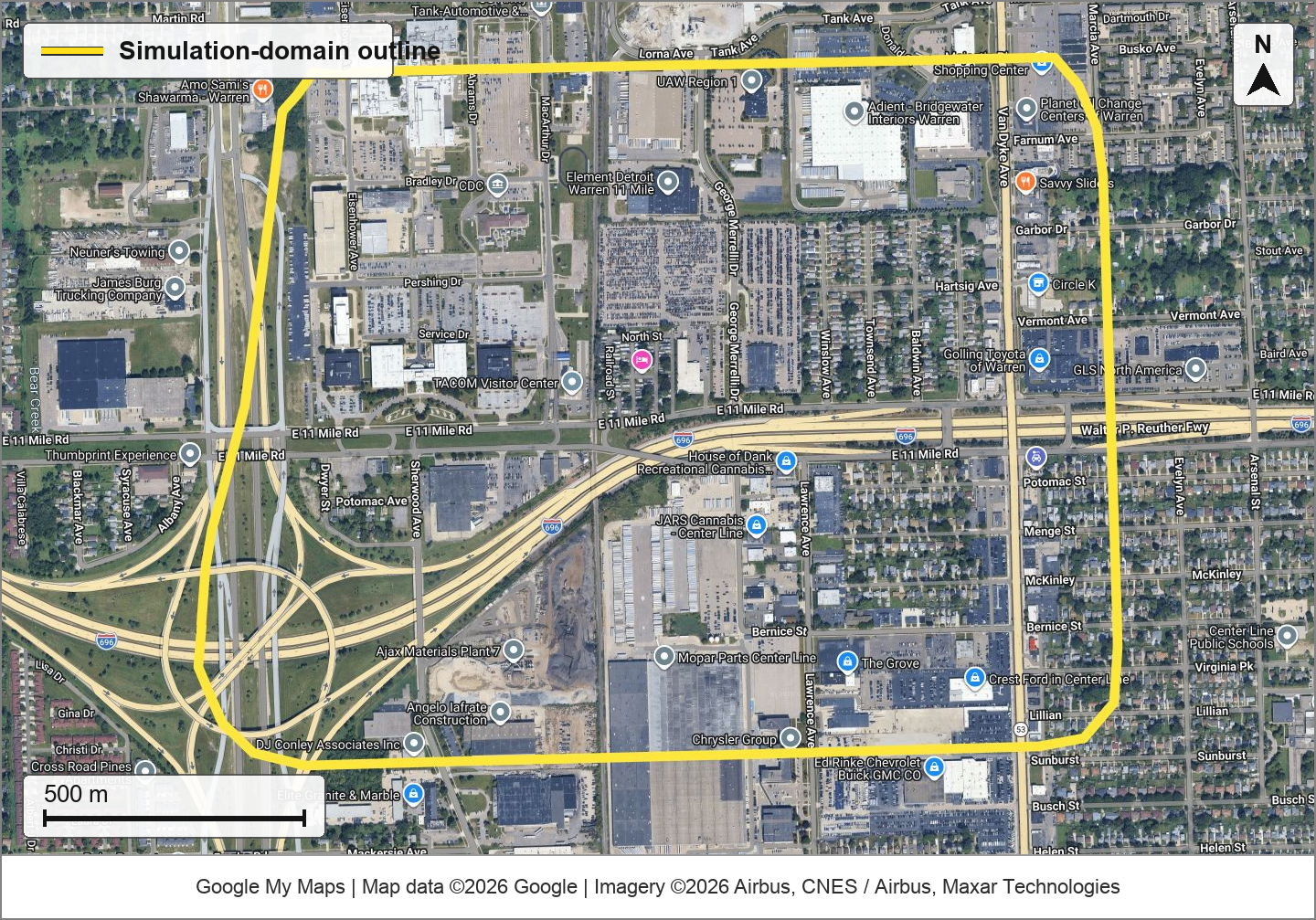}
    \caption{Present-day geographic context for the simulation domain (yellow outline) in the I-696--Mound Road corridor. The Google My Maps basemap is not part of the model input. Map and imagery were accessed on August 2, 2026; attribution appears in the image.}
    \label{fig:study_domain}
\end{figure}

\begin{figure}[!htb]
    \centering
    \includegraphics[width=.62\textwidth]{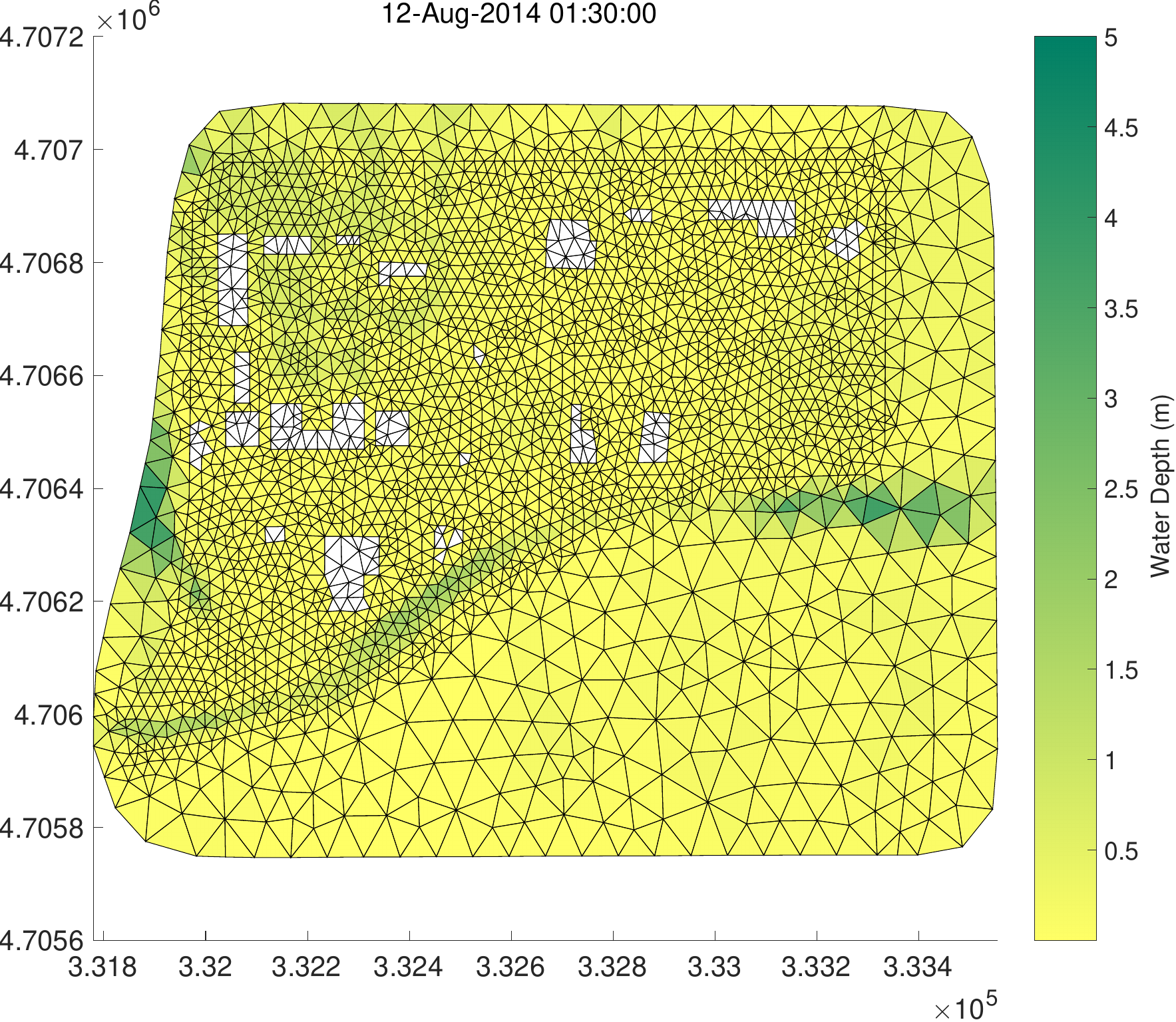}
    \caption{Simulated water depth in meters at 01:30 on 12 August 2014 for one tRIBS-Urban parameter realization.}
    \label{fig:snapshot}
\end{figure}

The case study is based on the August 2014 storm in the I-696--Mound Road corridor of metropolitan Detroit~\citep{FEMA2014}. The approximately $1.9\,\mathrm{km}^2$ Warren domain lies within the Red Run watershed and includes part of the Detroit Arsenal and Bear Creek~\citep{Tran2026}; \cref{fig:study_domain} provides present-day geographic context and outlines the simulation domain in yellow. The simulations use tRIBS-Urban to represent distributed hydrologic processes and two-dimensional overland flow across the urban terrain~\citep{Ivanov2004,Kim2012,Tran2024}. An unstructured, nonuniform triangular mesh constructed from $0.6\,\mathrm{m}$ LiDAR and building-footprint data contains 2,576 mesh nodes and 5,057 cells, with the greatest refinement northeast of the I-696--Mound Road intersection~\citep{Tran2026}. The model simulates infiltration, runoff generation, and overland flow from 05:00 on 11 August to 05:00 on 12 August 2014, with outputs available every 15 minutes; for a spatial location $x$, time $t$, and parameter realization $\param$, it returns the water depth $h(x,t;\param)$ in meters. \Cref{fig:snapshot} shows an example inundation field on the computational mesh.

The uncertain parameter random variable $\Param$ has six components describing antecedent soil moisture, infiltration, and soil-water retention. The case-study uncertainty specification assigns independent uniform priors,
\begin{align}
    \Param_i\overset{\mathrm{ind}}{\sim}
    \operatorname{Unif}(a_i,b_i),
    \qquad i=1,\ldots,6,
    \label{e:parameter_prior}
\end{align}
with the bounds shown in \cref{tab:parameters_version2}. The OED analysis conditions on the fixed precipitation forcing for the historical event and focuses on uncertainty in these six quantities, while topography, boundary conditions, spatial parameter variation, and the remaining model inputs stay at their case-study values.

\begin{table}[!htb]
    \centering
    \setlength{\tabcolsep}{3.5pt}
    \caption{Independent uniform priors used in the case study.}
    \label{tab:parameters_version2}
    \begin{tabular}{@{}cccl@{}}
        \toprule
        Component & Name & Uniform-prior bounds & Description \\
        \midrule
        $\Param_1$ & IniSM    & $[0.20,0.99]$       & Initial soil-moisture fraction $[-]$ \\
        $\Param_2$ & Ks       & $[8.99,9.82]$       & Saturated hydraulic conductivity $[\mathrm{mm\,h^{-1}}]$ \\
        $\Param_3$ & thetaS   & $[0.441,0.456]$     & Saturated soil moisture $[-]$ \\
        $\Param_4$ & thetaR   & $[0.115,0.124]$     & Residual soil moisture $[-]$ \\
        $\Param_5$ & m        & $[1.312,1.349]$     & Pore-distribution index $[-]$ \\
        $\Param_6$ & PsiB     & $[-0.007,-0.006]$   & Air-entry bubbling pressure head $[\mathrm{m}]$ \\
        \bottomrule
    \end{tabular}
\end{table}

\subsection{Surrogate model}

The design calculations require many flood-model evaluations across parameter realizations and candidate locations, making direct use of tRIBS-Urban computationally expensive at the scale required here. We therefore use a deep neural-network surrogate,
\begin{align}
    \widehat h(x,t;\param)
    \approx h(x,t;\param).
    \label{e:surrogate_model}
\end{align}
We train the surrogate using 200 tRIBS-Urban simulations sampled over the parameter ranges in \cref{tab:parameters_version2}. \Cref{app:surrogate} documents the network architecture and training procedure and compares tRIBS-Urban and surrogate inundation fields at 20:00 and 23:45 on 11 August 2014, where the displayed root-mean-square errors are on the order of $10^{-2}\,\mathrm{m}$. All subsequent calculations use $\widehat h$ in place of tRIBS-Urban.

\subsection{Observation and candidate designs}

Within each placement study, the observation time $t_o$ is fixed and $\design\in\designset$ denotes one candidate sensor location. The design set contains 2,576 locations corresponding to the simulation-mesh nodes, and the PO-OED analysis repeats the spatial ranking at several choices of $t_o$ to show how the preferred location changes over the event. At each location, the prospective observation follows
\begin{align}
    \begin{aligned}
        Y
        &=\widehat h(\design,t_o;\Param)+\mathcal{E},\\
        \mathcal{E}
        &\sim\mathcal{N}(0,\sigma_\epsilon^2),
          \qquad \mathcal{E}\perp\Param,
          \qquad \sigma_\epsilon=0.05\,\mathrm{m}.
    \end{aligned}
    \label{e:obs_model_prob}
\end{align}
The random variable $\mathcal{E}$ represents additive observation error. For this comparative study, an independent Gaussian distribution with $\sigma_\epsilon=0.05\,\mathrm{m}$ provides a common error model across all locations, while an operational design would select this value from the intended sensor and field conditions; the OED calculations treat the surrogate depth as the modeled response and apply this error model at every location.

\subsection{Prediction targets}

The prediction random variable is a deterministic function of the uncertain parameters,
\begin{align}
    Z=H(\Param).
    \label{e:pred_model_prob}
\end{align}
The distribution of $Z$ is induced by propagating the distribution of $\Param$ through $H$. We consider three point depths one hour after observation, their joint vector, regional-average depths, and vectors of regional maximum depth at three lead times, thereby representing monitoring purposes that range from a local forecast to a spatial or multitime summary. Within each study, the mapping $H$ is fixed before candidate locations are ranked and incorporates the surrogate, observation time, target geometry, lead times, and aggregation rule. Let $x_r$, for $r=1,2,3$, denote three chosen reference locations, let $\mathcal{R}_r$ contain the mesh nodes in target region $r$, and let $\mathcal{T}=\{1,2,5\}$ denote lead times in hours. The targets are
\begin{align}
    Z_{\mathrm{point},r}
    &=\widehat h(x_r,t_o+1\,\mathrm{h};\Param), \label{e:qoi_point}\\
    Z_{\mathrm{points}}
    &=\left(
      Z_{\mathrm{point},1},
      Z_{\mathrm{point},2},
      Z_{\mathrm{point},3}
      \right), \label{e:qoi_points}\\
    Z_{\mathrm{avg},r}
    &=\sum_{x_k\in\mathcal{R}_r}
      a_{rk} \widehat h(x_k,t_o+1\,\mathrm{h};\Param),
      \qquad
      a_{rk}\geq0,
      \qquad
      \sum_{x_k\in\mathcal{R}_r}a_{rk}=1, \label{e:qoi_average}\\
    Z_{\mathrm{max},r}
    &=\left(
      \max_{x_k\in\mathcal{R}_r}
      \widehat h(x_k,t_o+\tau\,\mathrm{h};\Param)
      \right)_{\tau\in\mathcal{T}}.
      \label{e:qoi_maximum}
\end{align}
The coefficients $a_{rk}$ define how water depths are averaged across the nonuniform mesh and are distinct from the stakeholder-priority weights introduced later: equal-node weights form an arithmetic mean over mesh nodes, while weights proportional to nodal control area approximate a physical area mean. Each component of $Z_{\mathrm{max},r}$ is the greatest modeled water depth in region $r$ at one specified lead time. The point locations and target regions are chosen to isolate how the learning target affects placement; an operational study would define them from the relevant warning threshold, asset, population, or response decision.

\begin{center}
\begin{minipage}{\textwidth}
    \centering
    \footnotesize
    \captionsetup{hypcap=false}
    \captionof{table}{Information targets used in the placement studies.}
    \label{tab:design_studies}
    \begin{tabularx}{\textwidth}{@{}
        >{\raggedright\arraybackslash}p{0.17\textwidth}
        >{\raggedright\arraybackslash}X
        >{\raggedright\arraybackslash}p{0.15\textwidth}
        >{\raggedright\arraybackslash}p{0.22\textwidth}@{}}
        \toprule
        Study & Information target & Prediction lead time & Criterion \\
        \midrule
        Parameter learning
        & Full six-component random variable $\Param$
        & ---
        & Joint parameter EIG \\
        Point prediction
        & $Z_{\mathrm{point},r}$ for each reference location
        & 1 hour
        & Point-prediction EIG \\
        Multiple points
        & Joint vector $Z_{\mathrm{points}}$ or its three marginals
        & 1 hour
        & Joint EIG or weighted sum of marginal EIGs \\
        Regional average
        & $Z_{\mathrm{avg},r}$ for each target region
        & 1 hour
        & Regional-average EIG \\
        Regional maxima
        & Vector $Z_{\mathrm{max},r}$ for each target region
        & 1, 2, and 5 hours
        & Joint prediction EIG \\
        \bottomrule
    \end{tabularx}
\end{minipage}
\end{center}

%% file: sections/03_procedure.tex
\section{Sensitivity analysis and Bayesian design criteria}\label{sec:procedure}

The analysis begins by examining how the six uncertain parameters influence modeled water depth, which provides physical context for the sensor rankings that follow. We then define two Bayesian design criteria: parameter EIG values measurements that improve knowledge of the full parameter vector, whereas prediction EIG values measurements that improve knowledge of a specified flood predictive QoI. The final subsection explains how these information measures connect with practical evidence used during deployment review.

\subsection{Global sensitivity analysis}

Global sensitivity analysis and OED answer complementary questions. Sobol' indices describe how uncertainty in each parameter contributes to variation in modeled water depth, whereas EIG describes how much information a prospective measurement is expected to provide about the selected learning target. A parameter may strongly affect the forward model without being readily learned from one noisy measurement, so the sensitivity analysis provides context for interpreting the information-based rankings rather than replacing them.

We use first- and total-order Sobol' indices to summarize parameter influence~\citep{Saltelli2008,Sobol2003}. The first-order index measures the fraction of water-depth variance attributable to one parameter acting alone, while the total-order index also includes every interaction involving that parameter, so the difference between them summarizes its interaction effects. The estimates use quasi-Monte Carlo samples from the independent priors and are evaluated at selected locations; \cref{app:sobol} provides the variance decomposition and sampling details.

\subsection{Parameter-oriented optimal experimental design}

PO-OED values a prospective observation by the information it provides about the full parameter vector $\Param$. Throughout this section, $p$ denotes a density whose arguments and conditioning identify the corresponding distribution. Before data collection, $p(\param)$ describes prior parameter uncertainty and the observation model in \eqref{e:obs_model_prob} supplies the likelihood $p(y|\param,\design)$. Once a measurement $y$ is collected at location $\design$, Bayes' rule combines them to give the posterior density,
\begin{align}
    p(\param|y,\design)
    =\frac{p(y|\param,\design)p(\param)}{p(y|\design)},
    \label{e:Bayes}
\end{align}
where the prior-predictive density $p(y|\design)$ acts as a normalization constant.

The Kullback--Leibler divergence measures how strongly this posterior differs from the prior for a given observation. Because the location must be selected before the measurement is known, parameter EIG averages that change over all prospective observations,
\begin{align}
    U_{\Param}(\design)
    &=\EE_{Y|\design}\!\left[
      \DKL\!\left(
      p(\param|Y,\design)\,\|\,p(\param)
      \right)\right]
      =\mathcal{I}(\Param;Y|\design).
    \label{e:EU_KL}
\end{align}
Natural logarithms express EIG in nats, and a larger value indicates that a measurement at that location is expected to produce a larger update in the parameter uncertainty. 
The PO-OED design is one that maximizes the parameter EIG:
\begin{align}
    \design_{\Param}^{*}
    \in\argmax_{\design\in\designset}U_{\Param}(\design).
    \label{e:PO-OED}
\end{align}

We numerically evaluate the criterion $U_{\Param}$ at every candidate location using nested Monte Carlo~\citep{Ryan2003}. \Cref{app:NMC} gives the estimator and sampling procedure.
We refer to the location with the largest computed value as the parameter max-EIG location.

\subsection{Goal-oriented optimal experimental design}\label{sec:GO_OED}

GO-OED changes the learning target from the full parameter vector to the flood predictive QoI $Z$. Propagating the prior and posterior parameter distributions through $H$ produces the prior-predictive density $p(z)$ and posterior-predictive density $p(z|y,\design)$, so prediction EIG measures the expected change between these two predictive distributions,
\begin{align}
    U_Z(\design)
    =\EE_{Y|\design}\!\left[
      \DKL\!\left(
      p(z|Y,\design)\,\|\,p(z)
      \right)\right]
      =\mathcal{I}(Z;Y|\design).
    \label{e:GO_EU_KL}
\end{align}
The GO-OED design is one that maximizes the prediction EIG:
\begin{align}
    \design_Z^{*}
    \in\argmax_{\design\in\designset}U_Z(\design).
    \label{e:GO-OED}
\end{align}
Because $Z$ is constructed from $\Param$, an observation cannot contain more information about $Z$ than it contains about the full parameter vector; nevertheless, the two criteria can rank locations differently because prediction EIG emphasizes only those parameter variations that affect the chosen prediction.

When $Z=(Z_1,\ldots,Z_m)$, the joint prediction EIG in \eqref{e:GO_EU_KL} values information about the predictive vector, including dependence among its components. A weighted marginal criterion instead evaluates each component separately and makes their relative priorities explicit,
\begin{align}
    U_Z^{w}(\design)
    =\sum_{j=1}^{m}w_j U_{Z_j}(\design),
    \qquad
    w_j\geq0,
    \qquad
    \sum_{j=1}^{m}w_j=1,
    \label{e:weighted_GO}
\end{align}
where $U_{Z_j}(\design)=\mathcal{I}(Z_j;Y|\design)$. When the components share information, the joint criterion accounts for that dependence directly, whereas the weighted criterion evaluates the contribution to each component and can emphasize the same shared information in several terms. If the prediction components are statistically independent both before and after the prospective observation, the joint criterion equals the unweighted sum of the marginal EIGs and therefore ranks locations like the equally weighted criterion; nearby flood depths generally share uncertain parameters and need not satisfy this condition.

Direct evaluation of the Kullback--Leibler expression in \eqref{e:GO_EU_KL} would require the posterior-predictive density for every possible observation and candidate location. We instead fit a conditional density model $q(z|y,\design;\lambda)$ using normalizing flows parameterized by $\lambda$, which can represent non-Gaussian predictive distributions~\citep{Dong2025}, and evaluate the variational criterion
\begin{align}
    U_L(\design;\lambda)
    &=\EE_{Y,Z|\design}\!\left[
      \log q(Z|Y,\design;\lambda)-\log p(Z)
      \right]
      \leq U_Z(\design).
    \label{e:GO_UL}
\end{align}
The bound becomes exact when the fitted conditional density matches the posterior-predictive density; therefore, we seek to maximize $U_L$ over both $\design$ and $\lambda$ simultaneously in order to get the tightest bound. Monte Carlo sampling is used to obtain a numerical estimate of this bound at each candidate location. While the second term in \eqref{e:GO_UL} does not depend on $\design$ or $\lambda$ and thus may be omitted from the optimization process without changing the optimizer, our calculations retain both terms so that $\widehat U_L$ represents the complete variational criterion rather than a relative ranking score. \Cref{app:variational} gives the estimator and fitting details. 

For readability, the remainder of the paper refers to the computed criteria as parameter EIG and prediction EIG, and to the location with the largest value in a given study as the max-EIG location. These terms refer to the numerical estimations described above.

\subsection{Integrating information value and deployment evidence}
\label{sec:proxy_classification}

EIG describes how informative a location is for the chosen learning target, while deployment evidence describes whether a sensor can be installed, operated, and maintained there. Verified restrictions on access, safety, permission, or infrastructure remove inadmissible locations from $\designset$ before ranking, whereas graded evidence about access, cost, power, telemetry, and maintenance can accompany a set of high-EIG locations during field review. Contextual features that define the desired flood prediction belong in $Z$ or its weights; features that affect installation and operation belong in the feasibility assessment.

The case study illustrates this workflow with public geospatial data from OpenStreetMap (OSM), an open geographic database distributed under the Open Database License~\citep{OpenStreetMapContributors2026a}. We select community-defined feature tags using Taginfo and query the model domain through the Overpass API~\citep{OpenStreetMapWikiContributors2026a,OpenStreetMapContributors2026b,OpenStreetMapWikiContributors2026b}. The queried features include roads, buildings, residential land use, power infrastructure, waterways, bridges, tunnels, access restrictions, barriers, military land use, and industrial or commercial land use.

The feasibility proxy gives its largest illustrative weight to proximity to an accessible road, uses buildings and power features as secondary indicators, and lowers the classification near mapped access restrictions, barriers, and military land. The context proxy combines building exposure, transportation, industrial or installation context, utility infrastructure, and hydrologic or drainage context, with the largest illustrative weights assigned to exposure and transportation. Context scores greater than $0.55$ are classified as high, scores from $0.25$ through $0.55$ as medium, and scores below $0.25$ as low; these thresholds organize the demonstration rather than define calibrated decision boundaries.

Because OSM represents conditions at its retrieval date, the resulting maps provide an initial organization of public evidence. An operational study would complete this assessment with local information about ownership, right-of-way, mounting, safety, power, telemetry, maintenance, sensor performance, and data governance, use verified constraints to define the admissible set before final ranking, and review its high-EIG locations using the remaining feasibility evidence.

%% file: sections/04_results.tex
\section{Results}\label{sec:results}

The results compare sensor placement under PO-OED with placement under GO-OED for several flood-prediction targets. Each EIG map ranks all 2,576 candidate locations at the stated observation time and reports values in nats; a black star marks the max-EIG location, while the posterior figures use synthetic observations to illustrate the corresponding uncertainty updates.

\subsection{Parameter contributions to water-depth variance}

\begin{figure}[!htb]
    \centering
    \includegraphics[width=\textwidth]{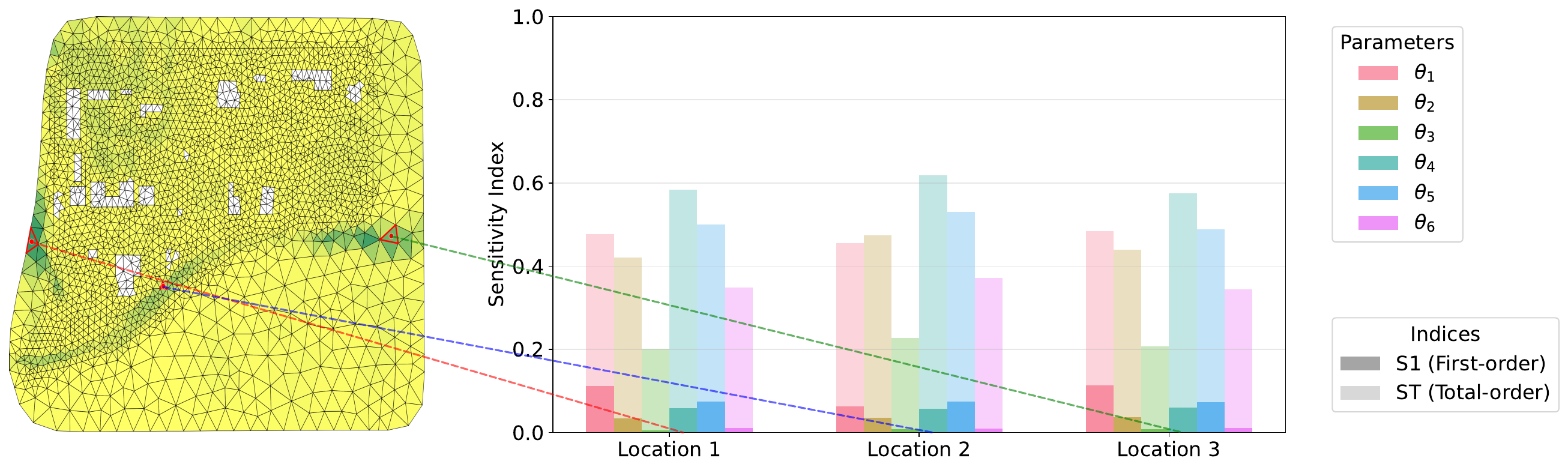}
    \caption{Estimated first-order ($S_i$) and total-order ($S_{T_i}$) Sobol' indices at three locations. Initial soil moisture has the largest first-order contribution at Locations 1 and 3, whereas the pore-distribution index has the largest contribution at Location 2. The gaps between first-order and total-order indices indicate parameter interactions.}
    \label{fig:sensitivity}
\end{figure}

\Cref{fig:sensitivity} shows the estimated first- and total-order Sobol' indices at three locations. Initial soil moisture, $\Param_1$, has the largest first-order contribution at Locations 1 and 3, whereas the pore-distribution index, $\Param_5$, has the largest contribution at Location 2, so the parameter with the strongest individual influence changes across the domain. At the examined locations, each estimated total-order index exceeds its first-order counterpart, indicating that interaction effects contribute to water-depth variance. This spatial variation and interaction motivate retaining the full six-parameter vector rather than fixing a subset.

\subsection{Sensor placement for parameter learning}

\begin{figure}[!htb]
    \centering
    \includegraphics[width=0.47\textwidth]{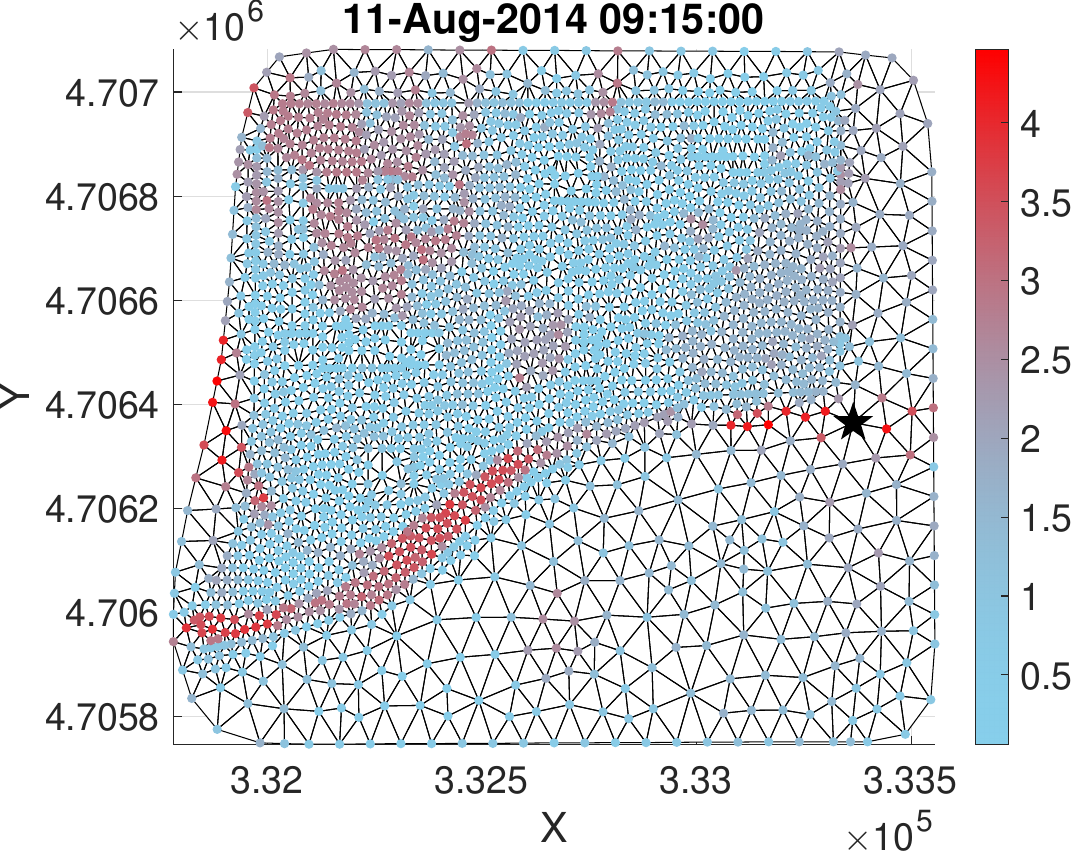}%
    \includegraphics[width=0.47\textwidth]{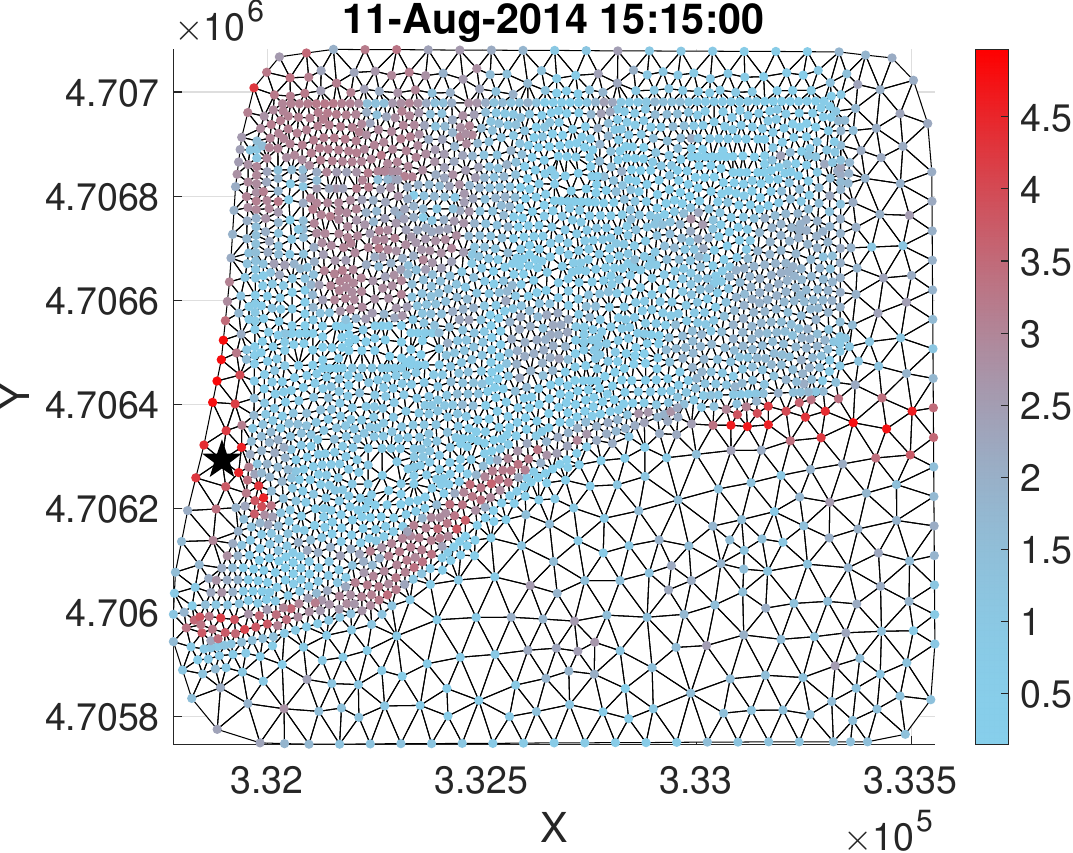}%
    \\
    \includegraphics[width=0.47\textwidth]{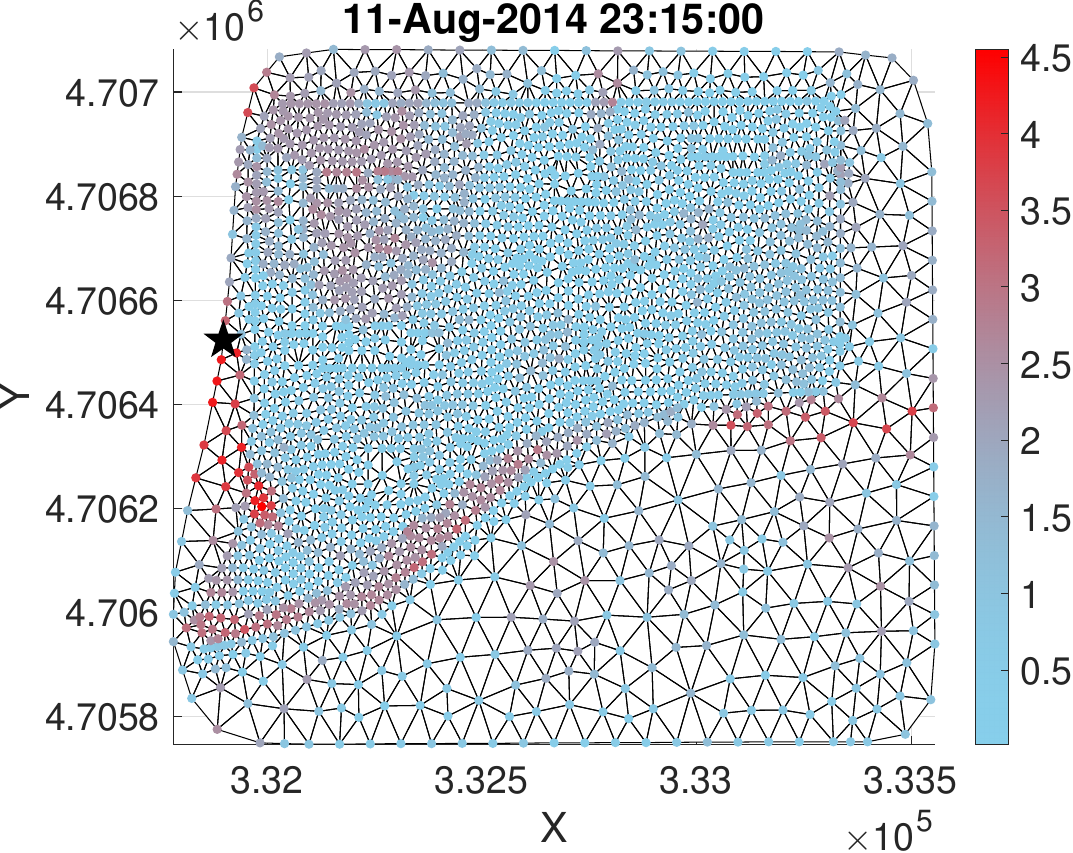}%
    \includegraphics[width=0.47\textwidth]{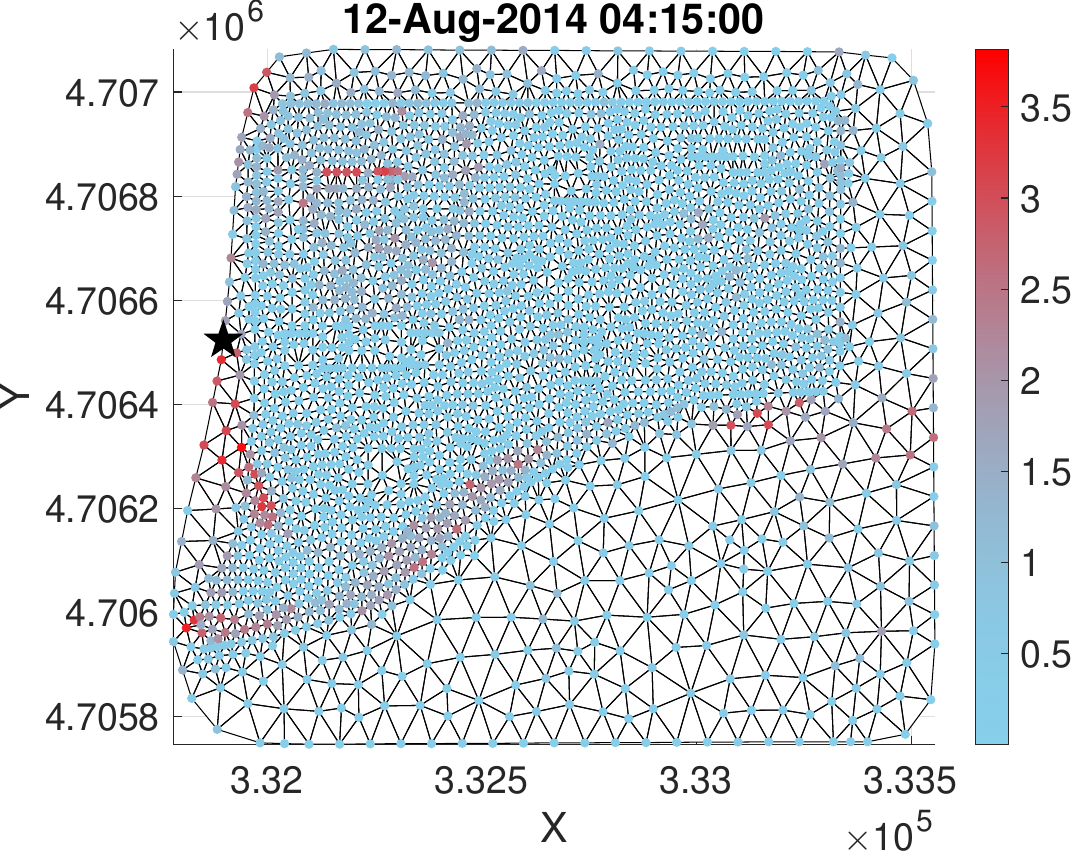}
    \caption{Parameter EIG at four observation times. From upper left to lower right, the maps correspond to 09:15, 15:15, and 23:15 on 11 August, followed by 04:15 on 12 August 2014. Colors show EIG in nats, and black stars mark the max-EIG locations.}
    \label{fig:OED_different_time}
\end{figure}

\Cref{fig:OED_different_time} shows that the spatial pattern of parameter EIG and the max-EIG location both change over the flood event, reflecting an evolving relationship between water depth and the uncertain parameters as the event develops. Observation time is therefore an important part of the monitoring design. Even when time and location are examined separately as in this case, a preferred location at one stage of the flood need not remain preferred later.

\begin{figure}[!htb]
    \centering
    \includegraphics[width=\textwidth]{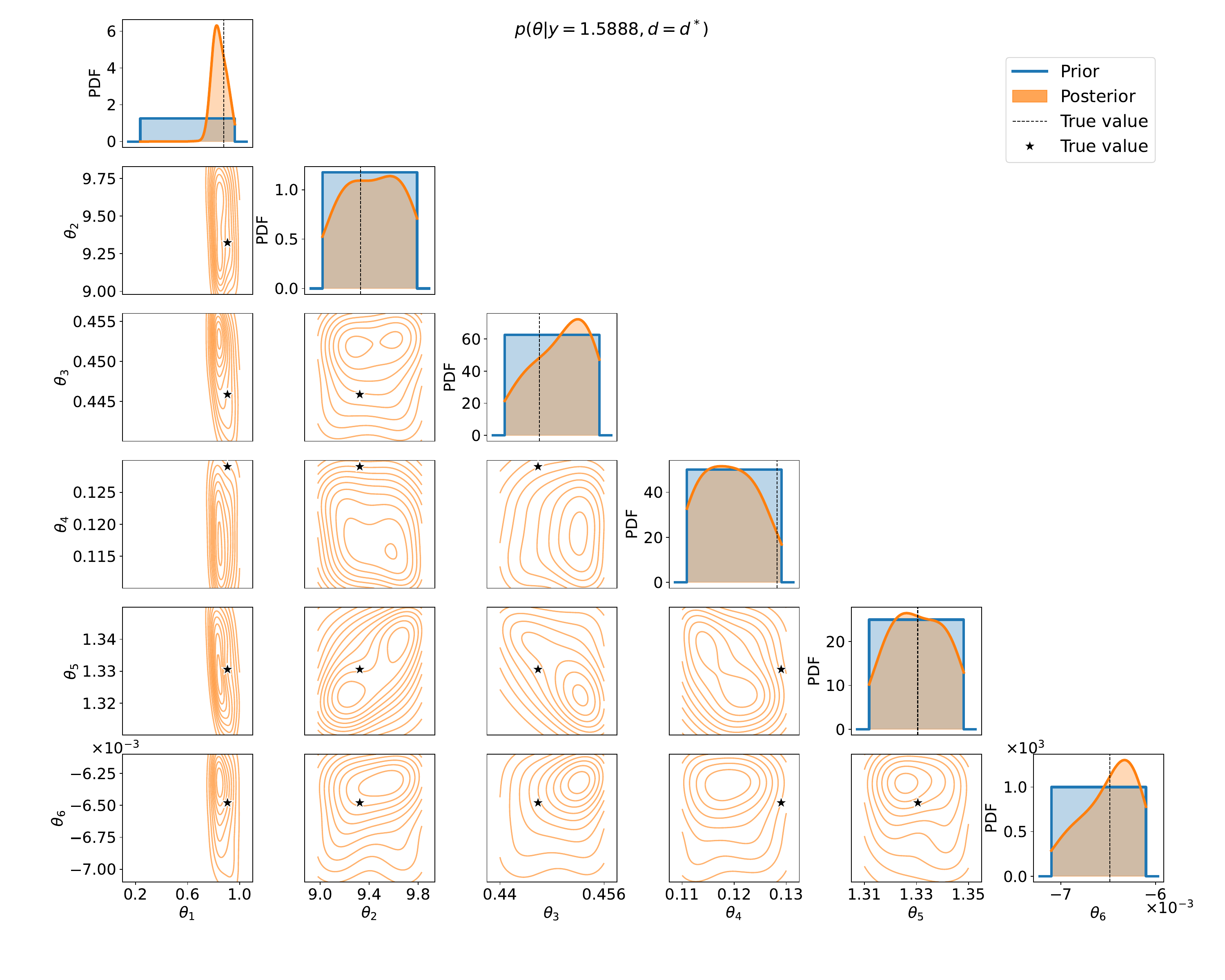}
    \caption{Parameter posteriors under a synthetic observation at 15:15 on 11 August 2014 at the parameter max-EIG location. Diagonal panels show marginal densities, lower panels show pairwise densities, and dashed lines and star markers denote the reference parameter values.}
    \label{fig:posterior_optimal}
\end{figure}

\begin{figure}[!htb]
    \centering
    \includegraphics[width=\textwidth]{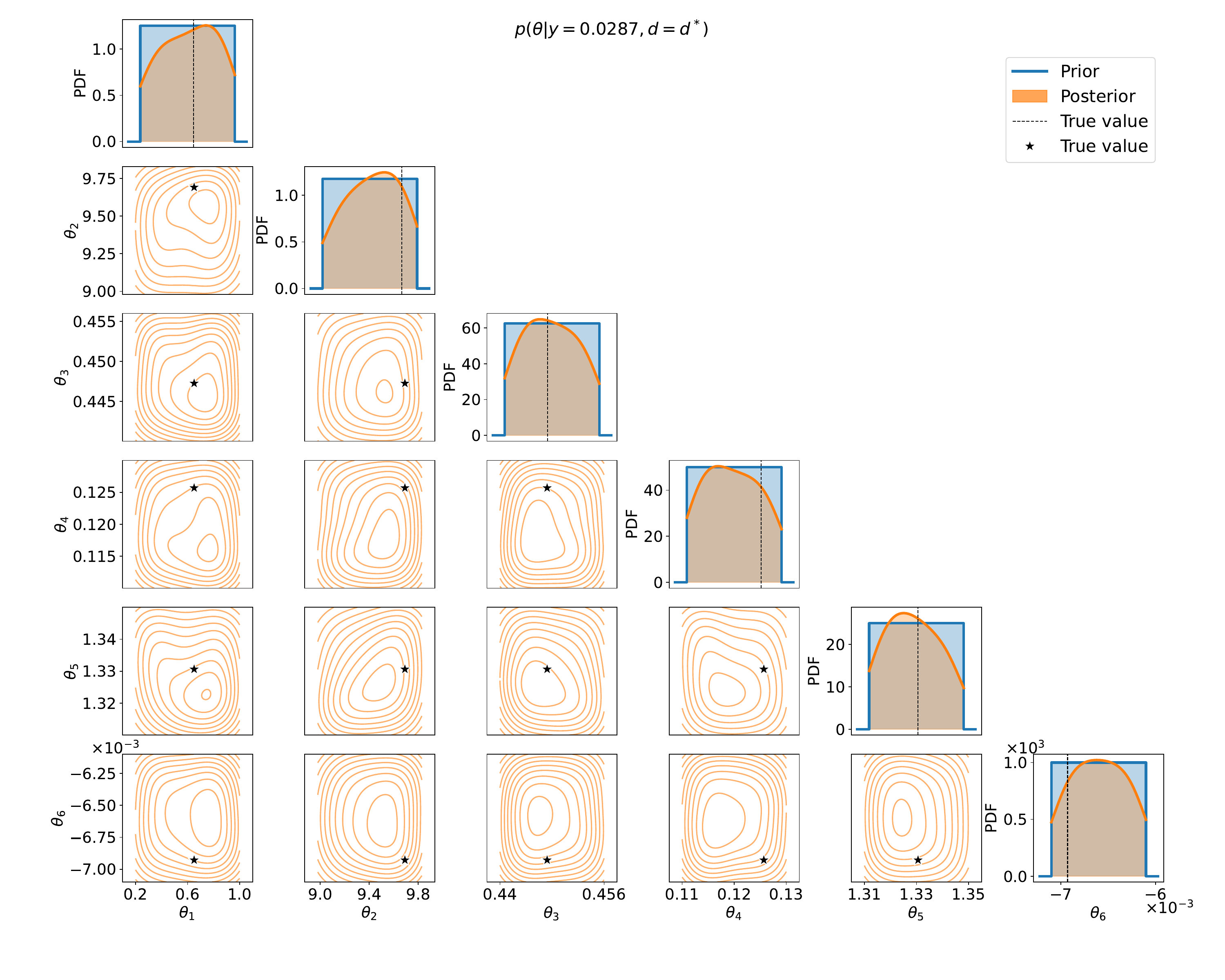}
    \caption{Parameter posteriors under a synthetic observation at 15:15 on 11 August 2014 at a location with small parameter EIG. Diagonal panels show marginal densities, lower panels show pairwise densities, and dashed lines and star markers denote the reference parameter values.}
    \label{fig:posterior_bad}
\end{figure}

\Cref{fig:posterior_optimal,fig:posterior_bad} compare marginal and pairwise parameter posteriors under synthetic observations at two candidate locations at 15:15 on 11 August 2014, generated from the same reference parameter realization. At the parameter max-EIG location (\cref{fig:posterior_optimal}), the posterior for initial soil moisture contracts strongly, while the other parameter marginals contract more moderately; at a location with small EIG (\cref{fig:posterior_bad}), the posteriors remain much closer to their priors. This comparison illustrates that the extent of uncertainty reduction can indeed be quite different for different location. Furthermore, the uncertainty reduction may be distributed unevenly across its six components.

\begin{figure}[!htb]
    \centering
    \includegraphics[width=\textwidth]{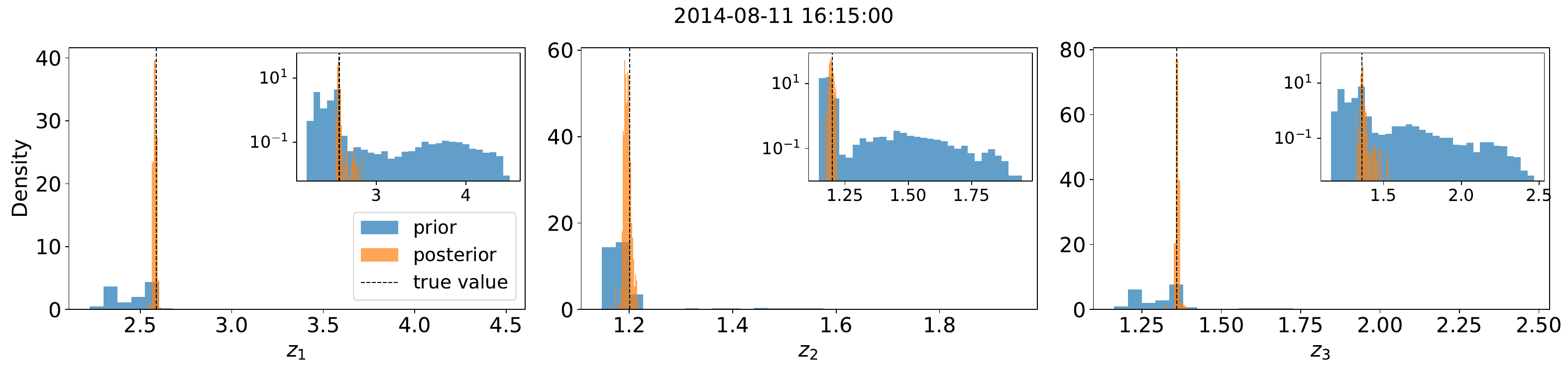}
    \includegraphics[width=\textwidth]{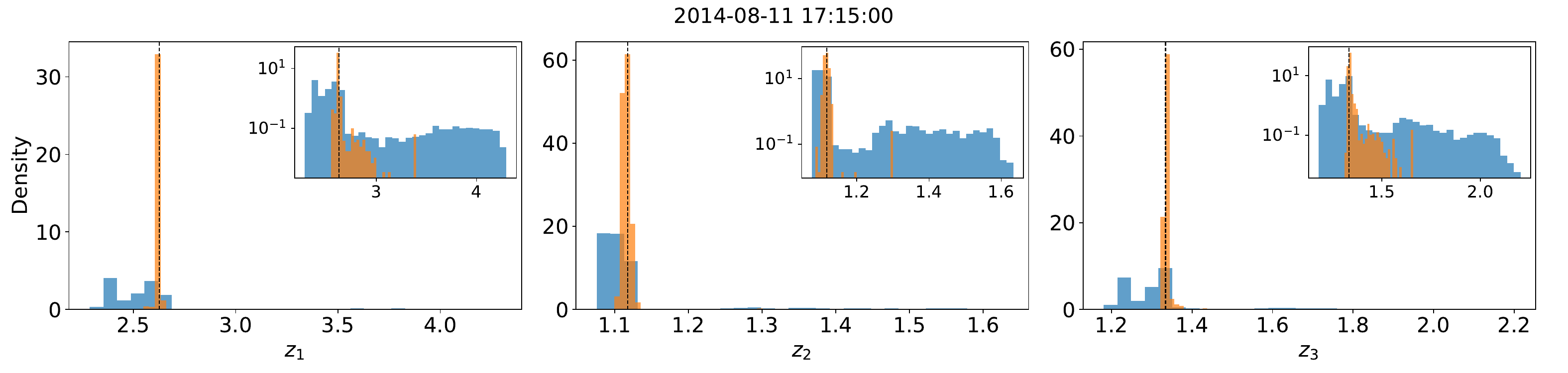}
    \includegraphics[width=\textwidth]{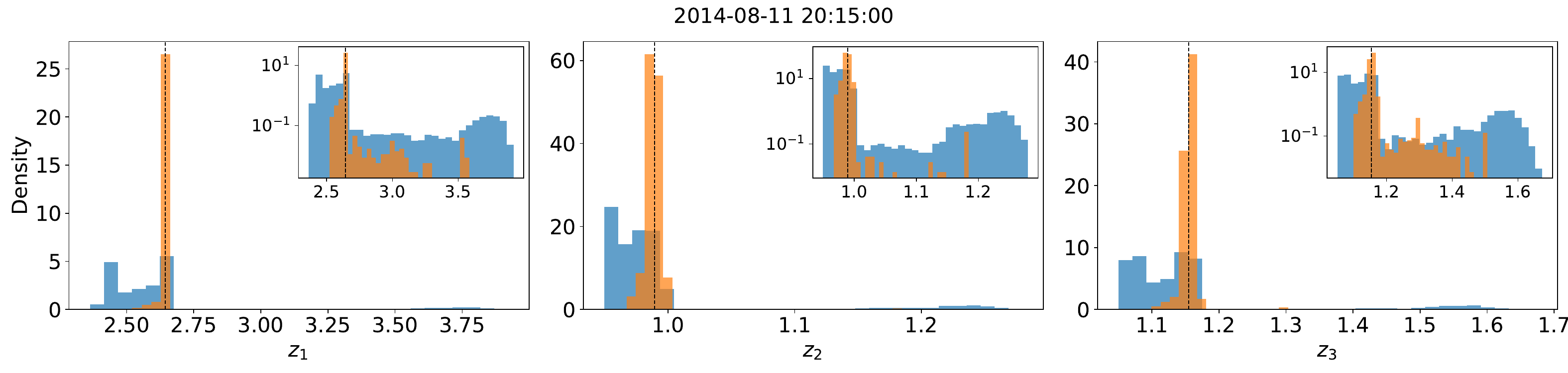}
    \caption{Posterior-predictive updates from a synthetic observation at 15:15 on 11 August 2014 at the parameter max-EIG location. Columns correspond to the three reference locations, and rows correspond to predictions 1, 2, and 5 hours after observation. Horizontal axes show water depth in meters; prior-predictive densities are blue, posterior-predictive densities are orange, insets use logarithmic density scales, and reference lines mark the simulated values.}
    \label{fig:posterior_predictive}
\end{figure}

\Cref{fig:posterior_predictive} propagates the parameter posterior from the max-EIG location at the 15:15 observation time to water-depth predictions at three locations and three lead times. The nine posterior-predictive distributions change by different amounts and in different ways, with some concentrating more sharply while others retain broader tails. Uncertainty reduction in parameters can therefore transfer unevenly to downstream predictions depending on the predictive QoI. This motivates choosing the prediction itself as the learning target when monitoring is intended to support a particular forecast or decision quantity.

\subsection{Sensor placement for flood prediction}

The GO-OED studies vary the prediction target $Z$ while holding the parameter prior, observation model, 15:15 observation time on 11 August 2014, candidate set, and surrogate fixed. This common setup allows us to separate out specifically the effect of how learning target changes sensor placement. 

\subsubsection{Point and multi-location prediction targets}

\begin{figure}[!htb]
    \centering
    \includegraphics[width=0.32\textwidth]{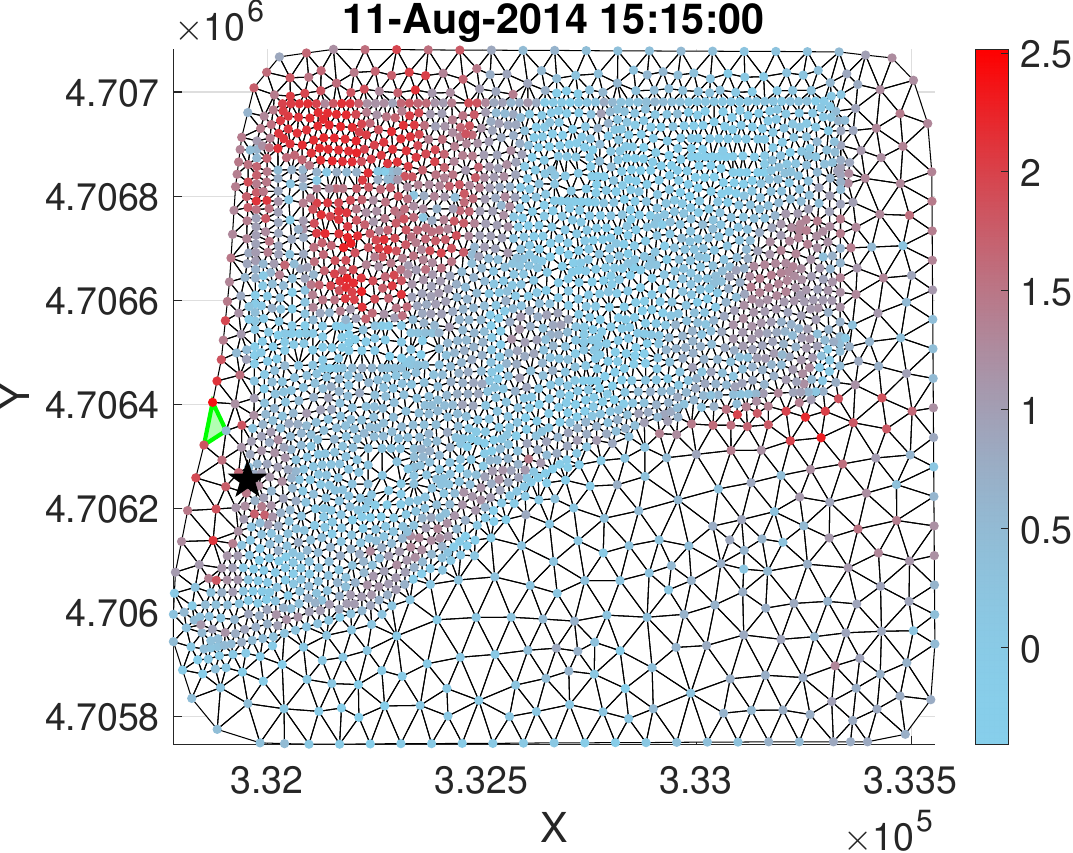}
    \includegraphics[width=0.32\textwidth]{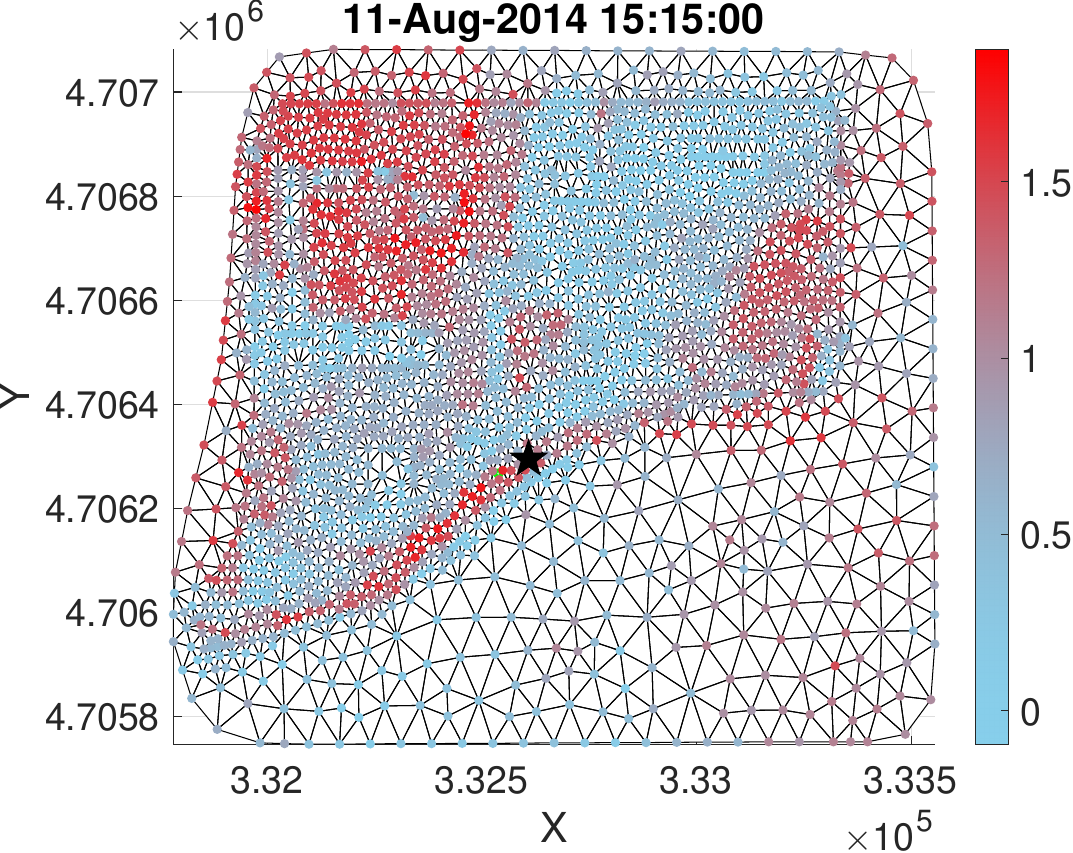}
    \includegraphics[width=0.32\textwidth]{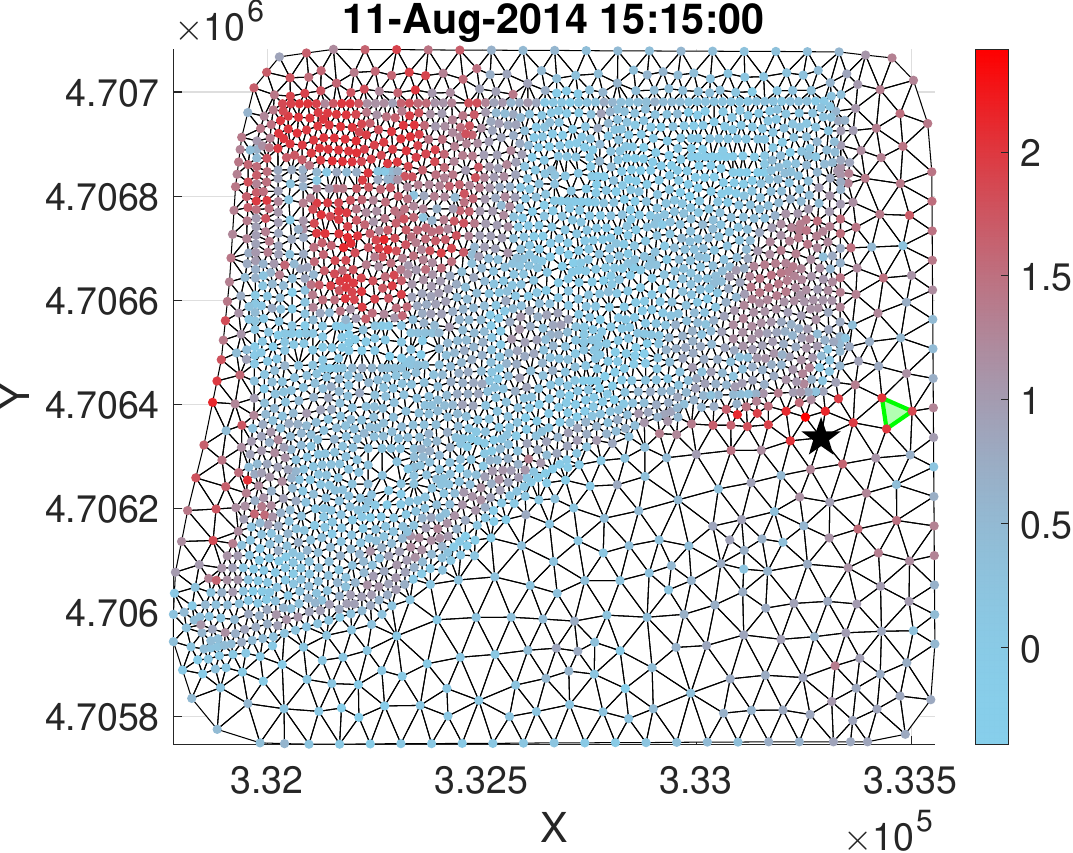}
    \caption{Prediction EIG for three one-hour point-depth targets, ordered from Location 1 to Location 3 from left to right. Green triangles identify the prediction targets, black stars mark the max-EIG locations, and colors show EIG in nats on target-specific scales.}
    \label{fig:GO_OED_single_loc_single_time}
\end{figure}

\Cref{fig:GO_OED_single_loc_single_time} shows prediction EIG for water depth one hour after observation at each of the three reference locations. The max-EIG location moves with the prediction target and lies nearby in all three cases, although the broader high-EIG regions extend beyond the target itself. For these point predictions, a nearby sensor is most informative under the working model, while the surrounding spatial pattern is consistent with information transmitted through shared parameter dependence and flood dynamics.

\FloatBarrier

\begin{figure}[!htb]
    \centering
    \subfloat[Joint target]{%
        \includegraphics[width=0.47\textwidth]{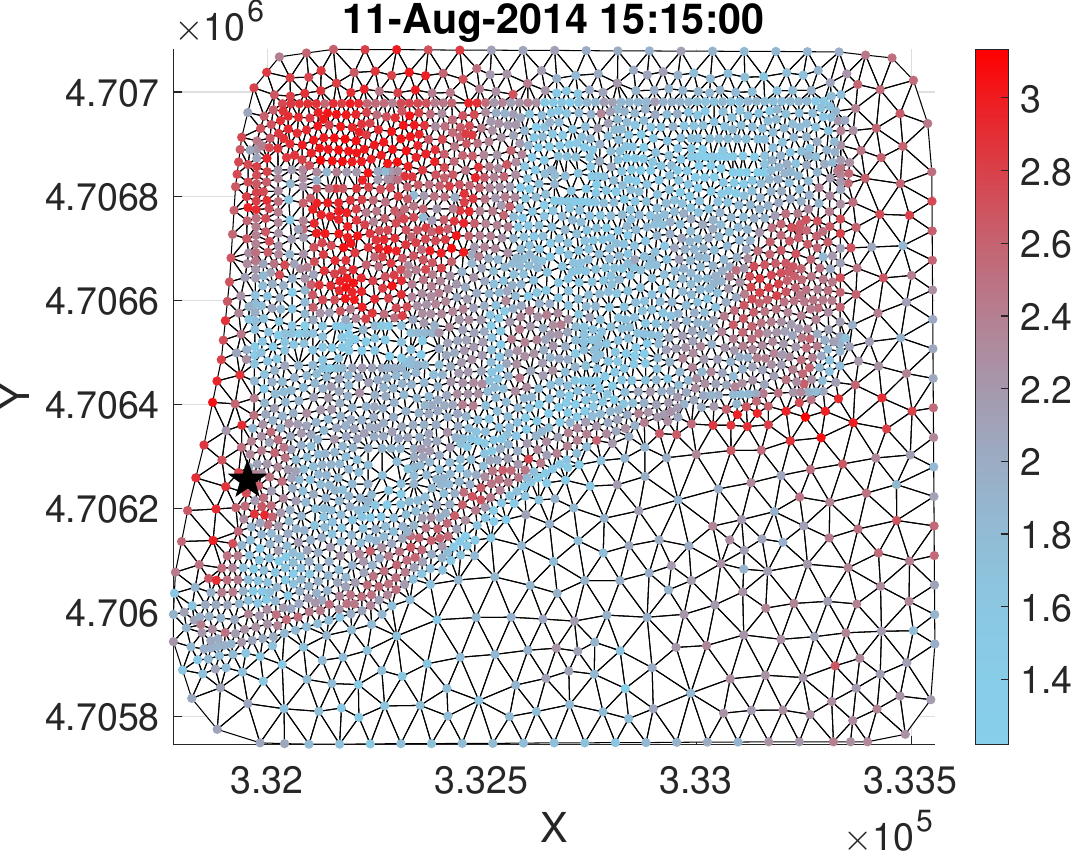}}%
    \subfloat[Equal weights]{%
        \includegraphics[width=0.47\textwidth]{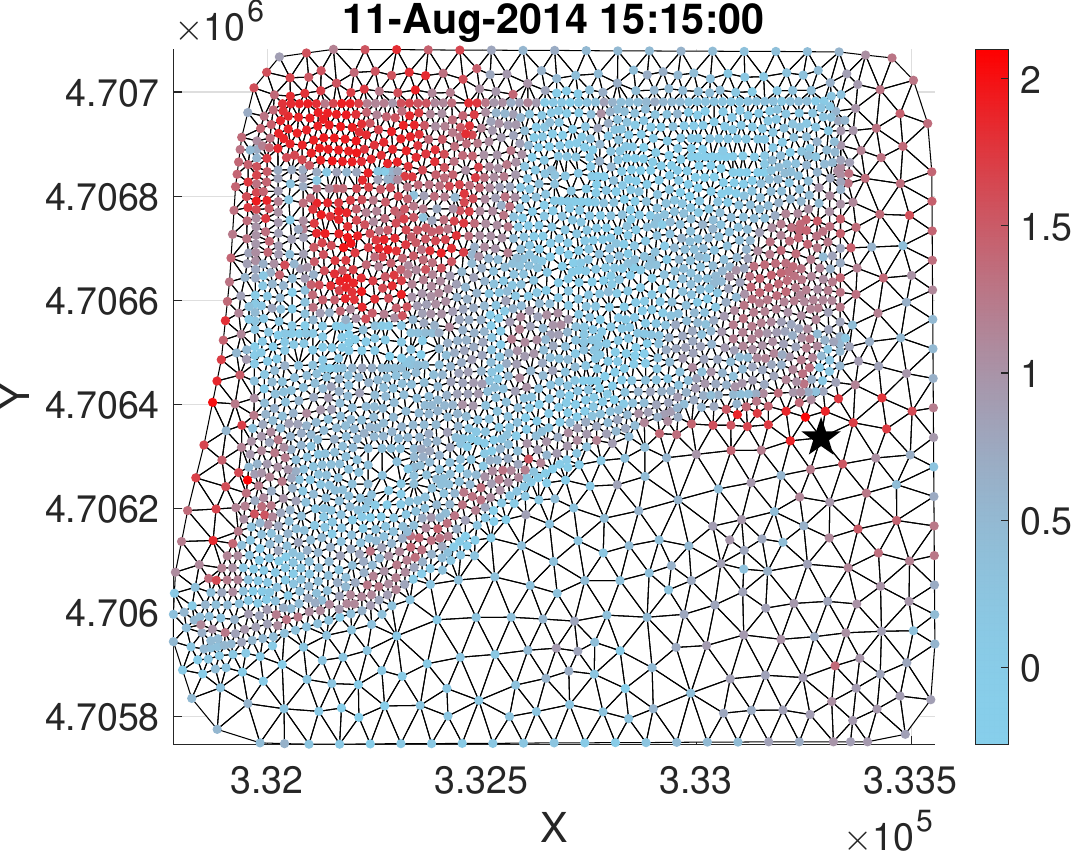}}%
    \\
    \subfloat[Scenario A]{%
        \includegraphics[width=0.47\textwidth]{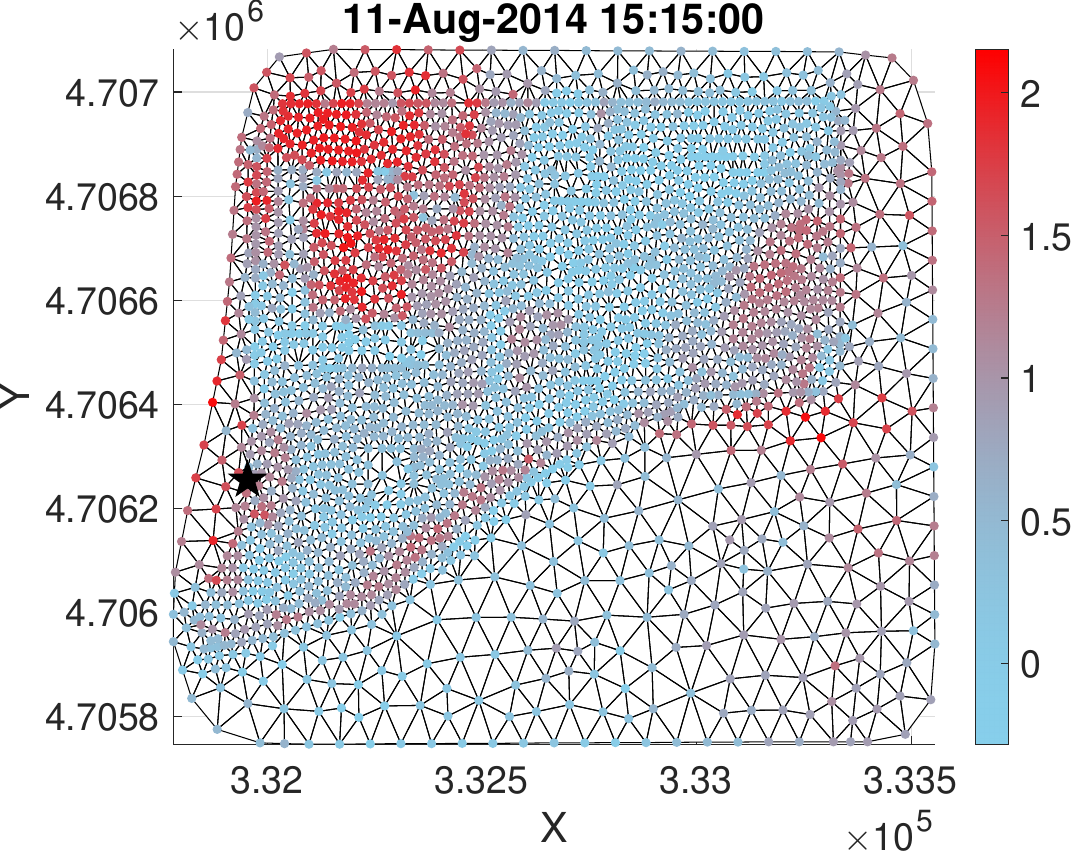}}%
    \subfloat[Scenario B]{%
        \includegraphics[width=0.47\textwidth]{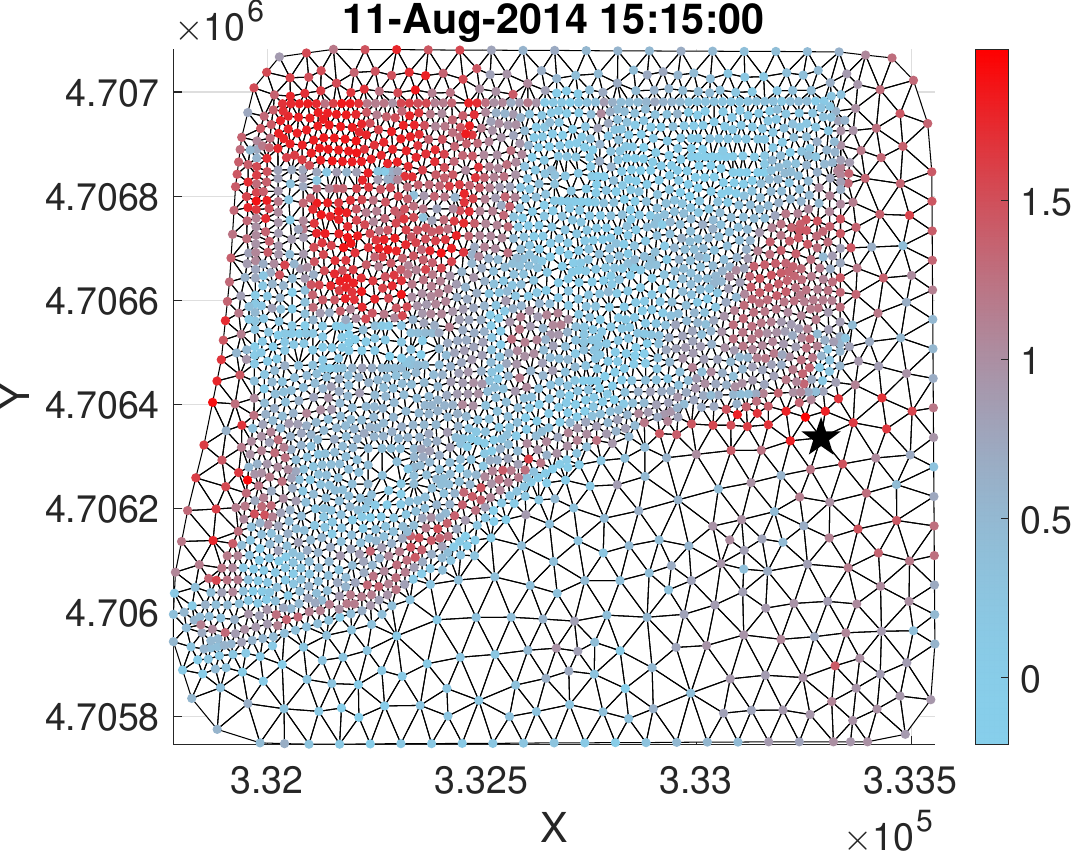}}
    \caption{Prediction EIG for a joint three-location target and three weighted marginal objectives. Green triangles identify the three targets, black stars mark the max-EIG locations, and colors show EIG in nats on panel-specific scales. The panels retain similar broad spatial patterns, while the scale and leading location depend on the objective.}
    \label{fig:GO_OED_multiple_locs_single_time}
\end{figure}

\clearpage
\Cref{fig:GO_OED_multiple_locs_single_time} compares a joint three-location target with equal marginal weights, Scenario A weights $(0.5,0.3,0.2)$, and Scenario B weights $(0.2,0.5,0.3)$. Despite their different numerical scales, the weighted cases retain similar broad patterns and recurring high-EIG regions. Equal weights and Scenario B produce an eastern max-EIG location, whereas the joint target and Scenario A produce a western one, but both locations remain prominent across the panels. This switch suggests close competition between two informative regions, which repeated calculations could assess more fully. \Cref{app:diagnostics} \cref{fig:GO_OED_posterior_multiple_locs_single_time} provides a synthetic-observation diagnostic for the joint target.

\FloatBarrier

\subsubsection{Regional-average prediction targets}

\begin{figure}[!htb]
    \centering
    \includegraphics[width=0.32\textwidth]{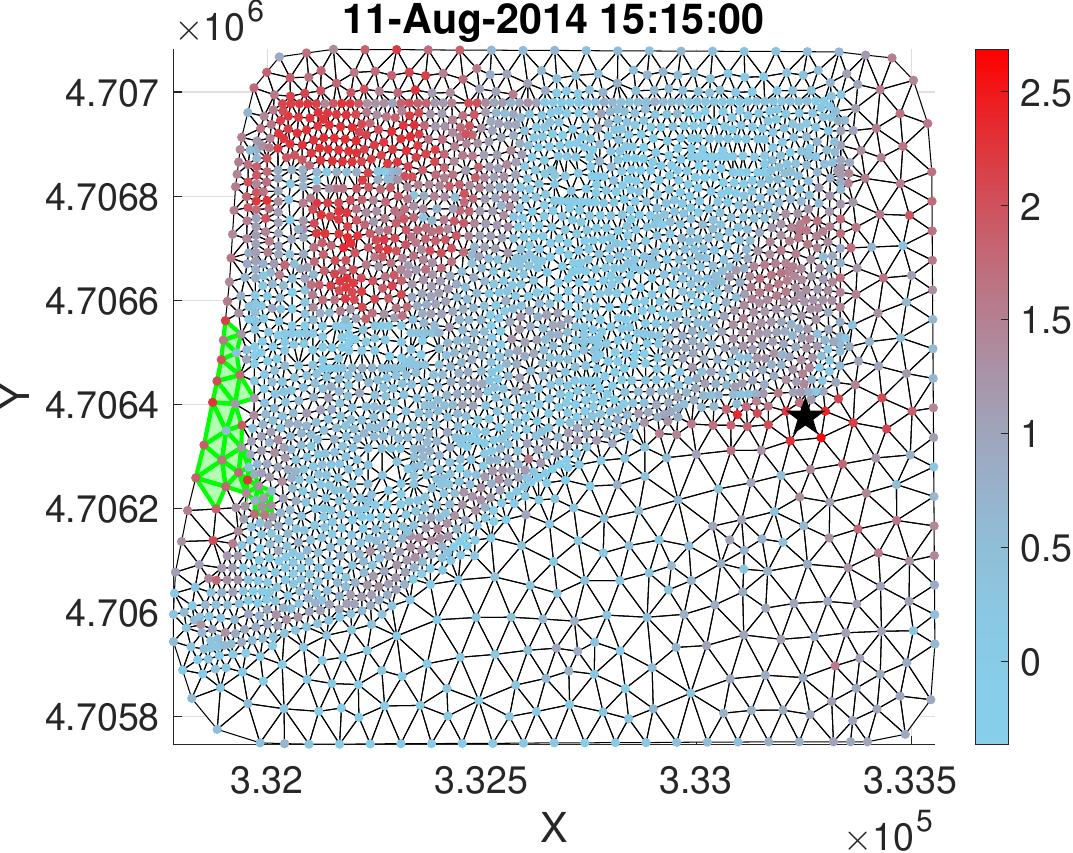}
    \includegraphics[width=0.32\textwidth]{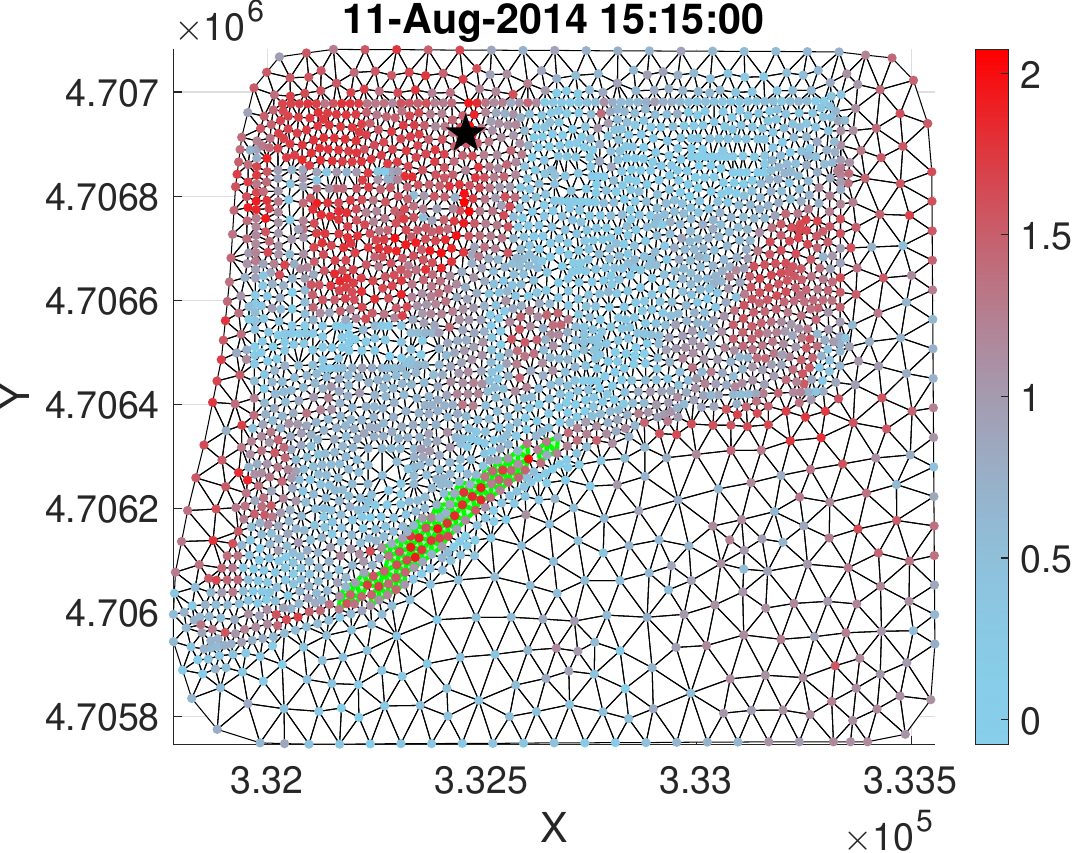}
    \includegraphics[width=0.32\textwidth]{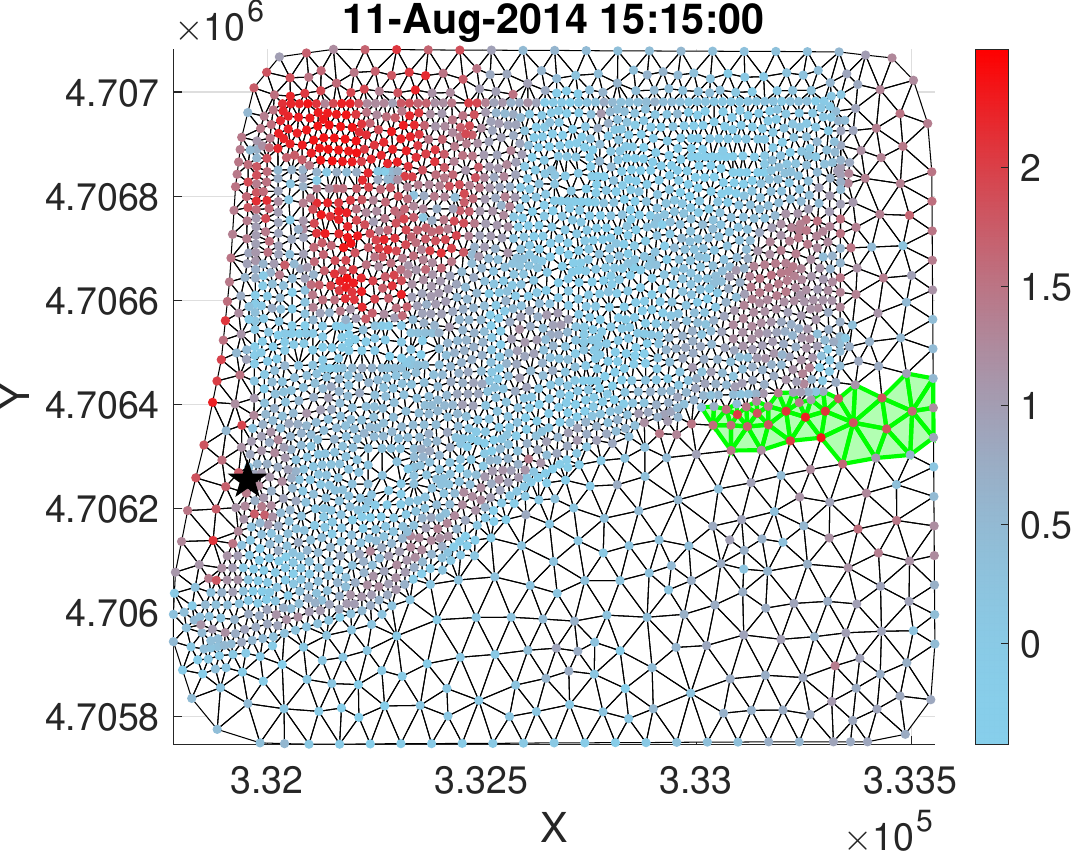}
    \caption{Prediction EIG for three one-hour regional-average targets, ordered from Region 1 to Region 3 from left to right. Green markers identify the target regions, black stars mark the max-EIG locations, and colors show EIG in nats on target-specific scales.}
    \label{fig:GO_OED_spatial_avg}
\end{figure}

\Cref{fig:GO_OED_spatial_avg} shows prediction EIG for average water depth one hour after observation in each target region. In all three cases, the max-EIG location lies outside the corresponding region, in contrast to the nearby placements obtained for point predictions. These nonlocal rankings indicate statistical dependence between a remote measurement and the regional-average prediction; hydraulic diagnostics can then examine whether shared parameter effects, flow pathways, or storage mechanisms explain that relationship. \Cref{app:diagnostics} \cref{fig:GO_OED_posterior_spatial_avg} provides a synthetic-observation diagnostic for the three regional averages.

\subsubsection{Regional maximum-depth prediction targets}

\begin{figure}
    \centering
    \includegraphics[width=0.32\textwidth]{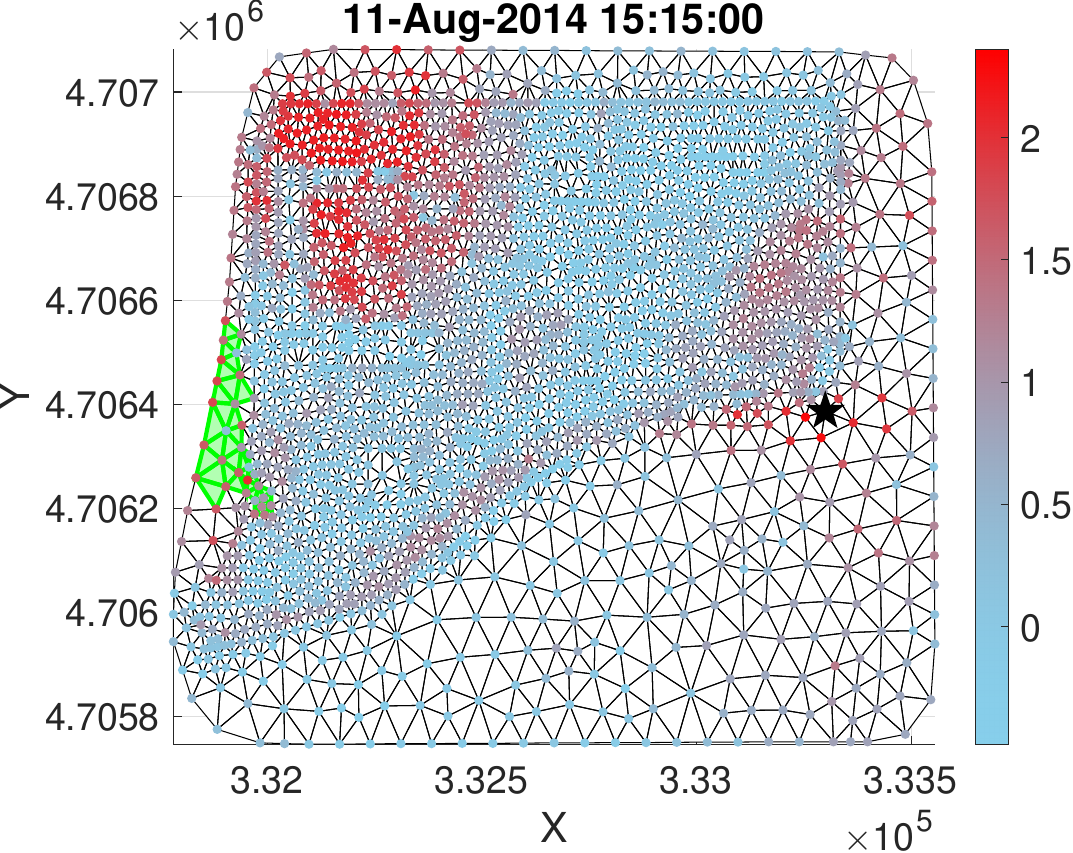}
    \includegraphics[width=0.32\textwidth]{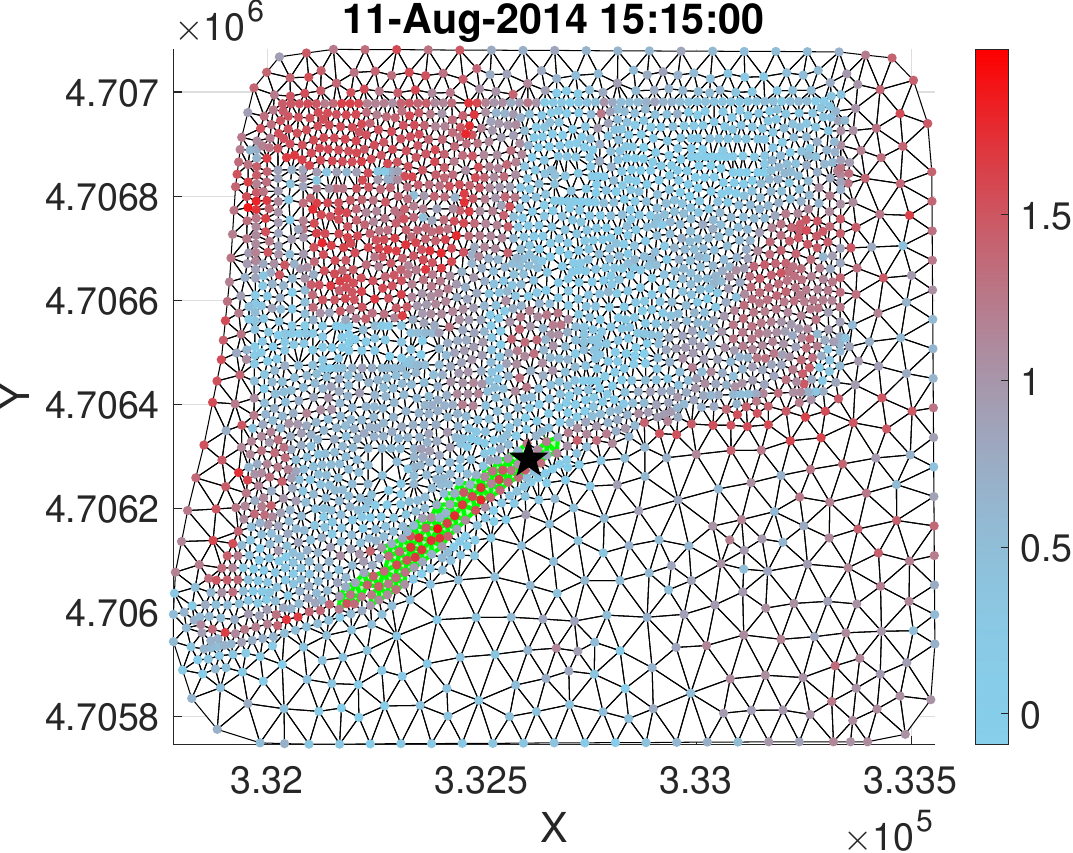}
    \includegraphics[width=0.32\textwidth]{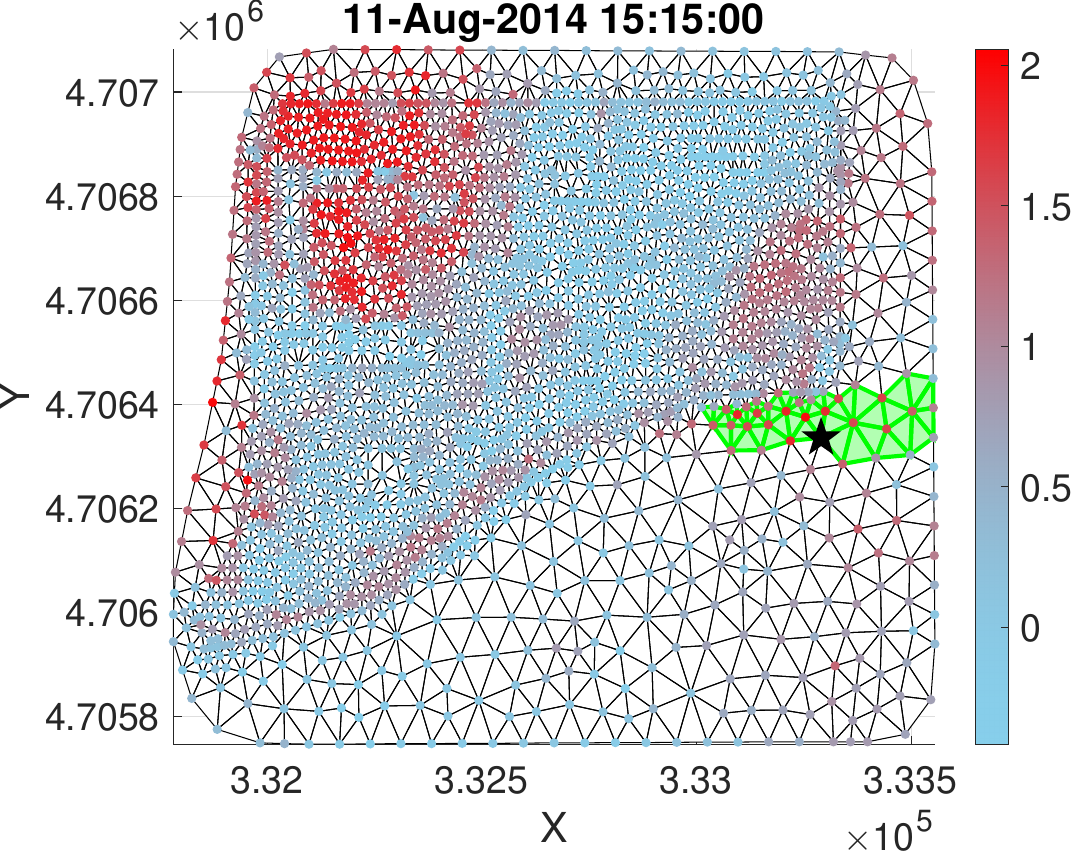}
    \caption{Prediction EIG for three regional maximum-depth vectors at lead times of 1, 2, and 5 hours, ordered from Region 1 to Region 3 from left to right. Green markers identify the target regions, black stars mark the max-EIG locations, and colors show EIG in nats on target-specific scales.}
    \label{fig:GO_OED_spatial_temporal_max}
\end{figure}

\Cref{fig:GO_OED_spatial_temporal_max} shows prediction EIG when each target is a vector of regional maximum depths at lead times of 1, 2, and 5 hours. The max-EIG location is remote from the first region, whereas the leading locations for the second and third regions lie close to or within their targets. Prediction-targeted placement can therefore be local or nonlocal, depending on the target region and how its evolving flood response is related to the rest of the domain.

Several GO-OED maps assign relatively large EIG to the northwestern part of the domain even when their maxima occur elsewhere. This recurring pattern identifies a useful region for hydraulic diagnosis of the shared parameter effects and flow pathways that connect prospective measurements to different prediction targets.

\vspace{1em}
Across the PO-OED and GO-OED studies, the computed max-EIG location changes with the learning target and observation time, although several broader high-EIG regions recur across cases. This target dependence suggests that a single heuristic placement rule (e.g., based on proximity or network coverage) is unlikely to be universally appropriate, and that Bayesian OED computations can provide quantitative, goal-tailored design optimizations.

\subsection{Deployment feasibility and context classifications}

\begin{figure}
    \centering
    \includegraphics[width=0.48\textwidth]{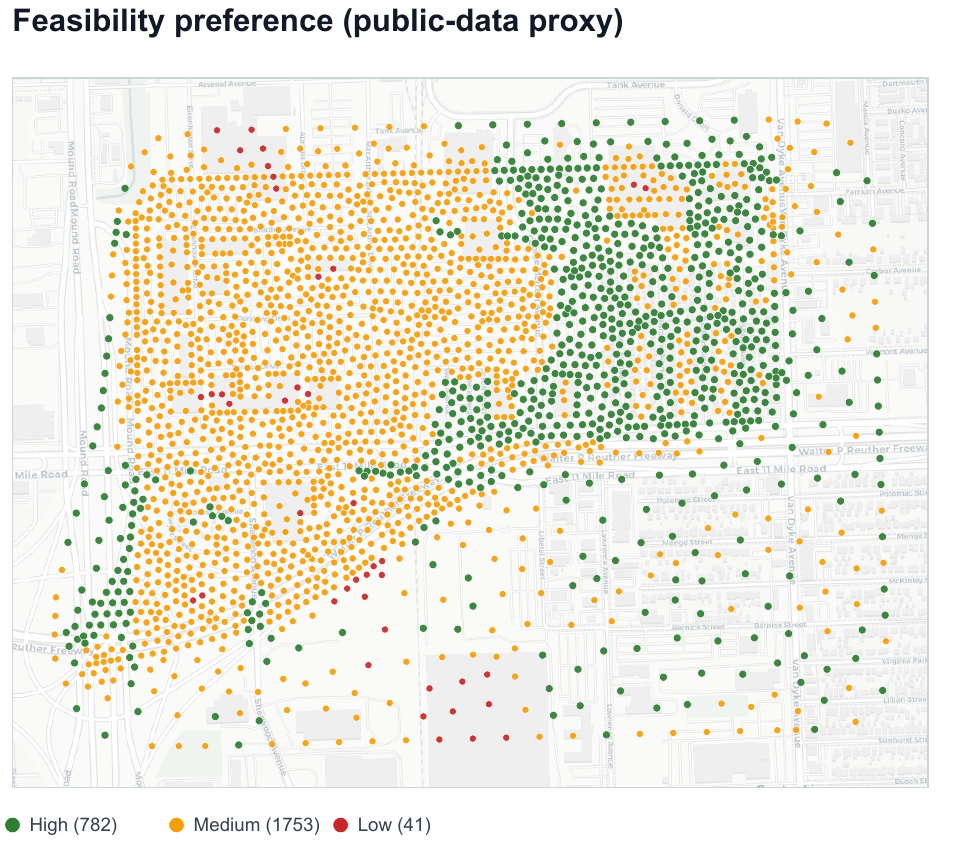}%
    \hfill
    \includegraphics[width=0.48\textwidth]{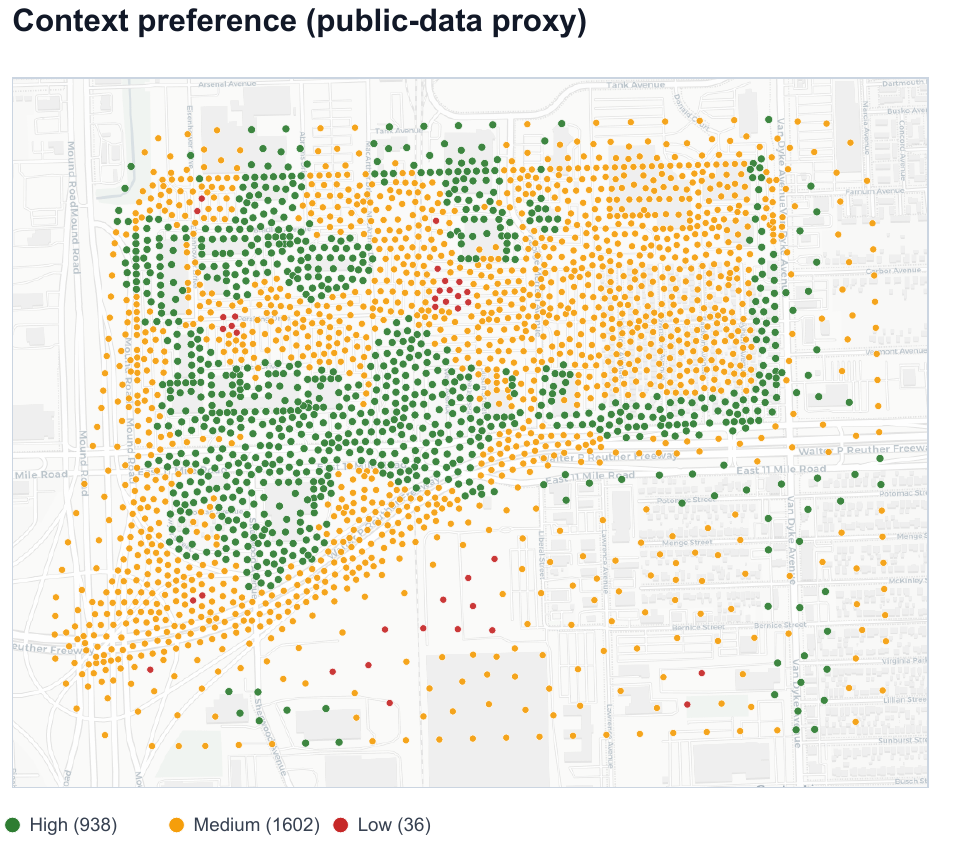}
    \caption{Illustrative feasibility and context classifications constructed from public geospatial data. Colors classify candidate locations as high (green), medium (orange), or low (red). The classes organize evidence for subsequent deployment review and require site-specific verification.}
    \label{fig:node_preference_map}
\end{figure}

\Cref{fig:node_preference_map} shows the feasibility and context classifications constructed using the procedure in \cref{sec:proxy_classification}. The feasibility proxy classifies 782 locations as high, 1,753 as medium, and 41 as low, while the context proxy classifies 938 locations as high, 1,602 as medium, and 36 as low. Their spatial patterns differ because feasibility emphasizes practical access for installation and maintenance, whereas context summarizes mapped exposure, infrastructure, and operational surroundings.

The two maps illustrate how deployment evidence can accompany an information-based ranking as a distinct design input. The present case study reports the classifications alongside the information maps to illustrate this screening stage, while an operational study would apply verified restrictions to the candidate set and use graded feasibility and context evidence to review the remaining high-EIG locations.

%% file: sections/05_discussion.tex
\section{Discussion and guidance for monitoring design}\label{sec:discussion}

The case study supports an objective-first approach in which the monitoring purpose is translated into a parameter or prediction target before candidate locations are ranked. The resulting information map is then interpreted with the flood model and considered together with evidence about deployment, connecting information value, physical meaning, and feasibility while preserving their distinct roles in the design process. \Cref{tab:practice_checklist} summarizes the questions that organize this workflow and the evidence provided by the current demonstration.

\begin{table}[!htb]
    \centering
    \footnotesize
    \setlength{\tabcolsep}{3pt}
    \caption{Reporting checklist and current evidence for information-based flood-sensor placement.}
    \label{tab:practice_checklist}
    \begin{tabularx}{\textwidth}{@{}
        >{\raggedright\arraybackslash}p{0.15\textwidth}
        >{\raggedright\arraybackslash}p{0.24\textwidth}
        >{\raggedright\arraybackslash}p{0.29\textwidth}
        >{\raggedright\arraybackslash}X@{}}
        \toprule
        Design choice & Question to resolve & Evidence to report & Current demonstration \\
        \midrule
        Monitoring purpose
        & Should data inform parameters or a predictive QoI?
        & State the criterion and connect it to the intended use of the data.
        & Parallel parameter- and prediction-EIG rankings illustrate objective dependence. \\
        Prediction target
        & Which location, region, horizon, statistic, or population defines the QoI?
        & Give $H$, spatial weights, lead times, and the source of priority weights.
        & Point, regional, and multitime targets with specified lead times. \\
        Numerical ranking
        & Which high-EIG locations remain competitive across repeated calculations?
        & Give sample sizes, repeated calculations, uncertainty summaries, and a high-EIG set.
        & One ranking per study; repeated calculations would quantify ranking stability. \\
        Physical interpretation
        & Why can the observation inform the target, especially when it is remote?
        & Compare predictive distributions and examine flow paths, storage, drainage, and shared parameter effects.
        & Synthetic-observation diagnostics motivate hydraulic analysis of nonlocal locations. \\
        Deployment review
        & Which sites are prohibited, feasible, costly, or operationally important?
        & Assign distinct roles to hard constraints, graded feasibility, and contextual evidence; verify public proxies locally.
        & Public-data classifications illustrate initial screening and require local verification. \\
        \bottomrule
    \end{tabularx}
\end{table}
\FloatBarrier

\subsection{Match the criterion to the monitoring purpose}

Parameter EIG and prediction EIG answer different monitoring questions, so the choice between them should follow the intended use of the data rather than a universal preference. PO-OED is appropriate when parameter learning is itself the purpose or when one posterior must support a range of predictions that have not yet been specified. At the parameter max-EIG location in the synthetic example, initial soil moisture exhibits the strongest marginal posterior contraction, while the resulting predictive changes vary across locations and lead times, which is consistent with the fact that no single prediction defines the criterion.

GO-OED is more direct when monitoring serves a named forecast or decision quantity, because prediction EIG measures the statistical dependence between a prospective observation and that quantity. For the point targets in this event, the max-EIG locations lie nearby, whereas regional-average and regional maximum-depth targets can favor remote locations. Prediction EIG therefore complements familiar considerations such as distance and modeled depth by identifying measurements that are informative for the specific intended prediction.

The variation in spatial rankings across monitoring objectives also shows why a single placement heuristic (e.g., based on proximity or network coverage) would not necessarily serve well across different purposes; rather, their relevance depends on the intended use of the data. Bayesian OED is able to provide a systematic way to identify high-value locations for each monitoring scenario.

\subsection{Define the prediction target before ranking locations}

Optimality is defined relative to the learning target, making target specification a substantive part of the monitoring-design problem. If the target does not represent the forecast or decision that the monitoring program is intended to support, additional computational effort can solve the stated design problem more accurately but cannot make the resulting design aligned with the actual monitoring need. The QoI should therefore specify its spatial support, lead time, threshold, aggregation rule, and represented asset or population before candidate locations are ranked.
On the nonuniform mesh used here, an arithmetic average over mesh nodes gives greater influence to finely resolved areas than an area-weighted average, while a vector of regional maxima at selected lead times answers a different question from the single largest depth over an entire space--time region.

A joint criterion values information about the predictions as a vector, including their dependence, whereas a weighted marginal criterion makes the relative priority of each prediction explicit and evaluates its information contribution separately. The case-study weights are sensitivity scenarios rather than elicited stakeholder preferences; an operational study would document their source, represent disagreement, and examine whether plausible weights preserve a common high-EIG region.

EIG measures the expected reduction in uncertainty across the full predictive distribution, so posterior-predictive diagnostics can examine contraction, shifts, tails, multimodality, and decision-relevant threshold probabilities rather than variance alone. This criterion is appropriate when distributional learning is the monitoring goal; a design aimed at a forecast score, threshold classification, or operational decision can instead use a utility aligned with that outcome.

\subsection{Interpret information maps as model-based evidence}

An information map conveys spatial structure beyond its largest value. Broad regions of similarly large EIG can provide placement flexibility, whereas an isolated maximum merits closer examination; repeated calculations, surrogate perturbations, prior-sensitivity studies, and alternative observation times can reveal which features of the ranking persist and which locations are interchangeable.

A remote measurement can be informative when it and the prediction target share sensitivity to the same uncertain parameters, although the statistical relationship alone does not identify the hydraulic mechanism. Correlations between prospective sensor measurements and the target prediction, targeted parameter perturbations, and flow-path analyses can test explanations based on routing, storage, drainage connectivity, topography, or built-surface representation. In the present case, the high-EIG region that recurs in the northwest across several targets provides a focused starting point for that diagnosis.

\subsection{Integrate information value with deployment evidence}

Deployment inputs enter the analysis according to their meaning. Verified prohibitions and safety restrictions define the feasible design set, while prediction priorities define $Z$ or its weights. Within the feasible set, graded evidence about access, cost, maintenance, power, and telemetry can support constrained optimization, explicit comparison of information value and deployment burden, or review of high-EIG locations. When exposure or infrastructure defines the QoI, it belongs to the monitoring purpose; it also belongs in deployment review when it creates an independent operational consideration.

Public OpenStreetMap features provide an initial way to organize feasibility and contextual evidence. A reproducible screening analysis records the queries, distance rules, thresholds, numerical weights, and processing code, while site visits and local review establish ownership, right-of-way, safety, mounting, power, telemetry, maintenance, sensor reliability, and data governance. The resulting assessment can then report the feasibility of each high-EIG location and the change in information value imposed by the verified feasible set, thereby completing the progression from information analysis to a deployment recommendation.

\subsection{Scope and opportunities for extension}

The event-specific analysis uses one urban domain, one historical precipitation forcing, and a single-sensor design problem, which creates a controlled setting for isolating how the monitoring objective changes placement. Applying the same workflow to uncertain future rainfall, additional flood events, sensor networks, adaptive schedules, or mobile observations would connect this comparison more directly to operational forecasting and broader monitoring strategies.

The six-parameter model and neural-network surrogate likewise provide a focused setting for the information calculations. Future analyses can test sensitivity to the prior and observation-error model, expand the uncertain inputs to precipitation, topography, boundary conditions, spatial parameter variation, observation-model misspecification, and model discrepancy, and validate the surrogate across the prior, wet and dry regimes, and high-EIG locations to establish whether it preserves the spatial rankings. Repeated nested Monte Carlo calculations and variational fits would complement this validation by identifying stable high-EIG regions and distinguishing close competitors, particularly for the weighted multi-point cases in \cref{fig:GO_OED_multiple_locs_single_time}.

%% file: sections/06_conclusion.tex
\section{Conclusions}\label{sec:conclusion}

This paper combines global sensitivity analysis with parameter EIG and prediction EIG for single-sensor placement in an urban flood-inundation case study. Its contribution lies in the controlled application and interpretation of established Bayesian OED methods, together with practical guidance for matching monitoring locations to their intended use. We extract three main conclusions.

\begin{enumerate}
    \item Parameter EIG and prediction EIG support different monitoring purposes. At the examined locations, the estimated Sobol' indices indicate interaction effects among the uncertain parameters, while the PO-OED example illustrates that learning the full parameter vector can transfer unevenly to predictions across locations and lead times.

    \item The prediction target shapes the spatial ranking. Point-depth targets favor nearby locations in this case, whereas regional averages and vectors of regional maximum depth can favor nonlocal locations. Weighted multi-point objectives preserve similar broad spatial patterns, with their computed max-EIG locations falling in one of two visually competitive regions across the priority scenarios.

    \item Information value and deployability belong in the same design process but require different evidence. Public geospatial classifications can organize initial screening, while verified field conditions determine which informative locations are feasible for installation and long-term operation.
\end{enumerate}

For practice, these findings support an objective-first workflow: define the intended prediction and any priority weights, use field-verified restrictions to establish the admissible set, compute the corresponding information map, identify a set of high-EIG locations, and interpret them using hydraulic and graded feasibility evidence. The present event and uncertainty model provide a focused demonstration of this workflow, while repeated ranking analyses, additional events, broader surrogate validation, stakeholder elicitation, and field assessment offer a path from the current comparison to deployable monitoring recommendations.

%% file: sections/07_appendix.tex
\appendix

\section{Surrogate model assessment}\label{app:surrogate}

The deep neural-network surrogate $\widehat h$ maps spatial location, time, and the parameter realization $\param$ to water depth. Its training set contains 200 tRIBS-Urban simulations sampled over the parameter ranges in \cref{tab:parameters_version2}. The network has five hidden layers with 256 neurons per layer, rectified linear-unit activations, and dropout regularization; training uses the Adam optimizer with learning rate $0.001$ and batch size $5{,}000$.

\begin{figure}[!htb]
    \centering
    \subfloat[11 August 2014, 20:00]{%
        \includegraphics[width=0.47\textwidth]{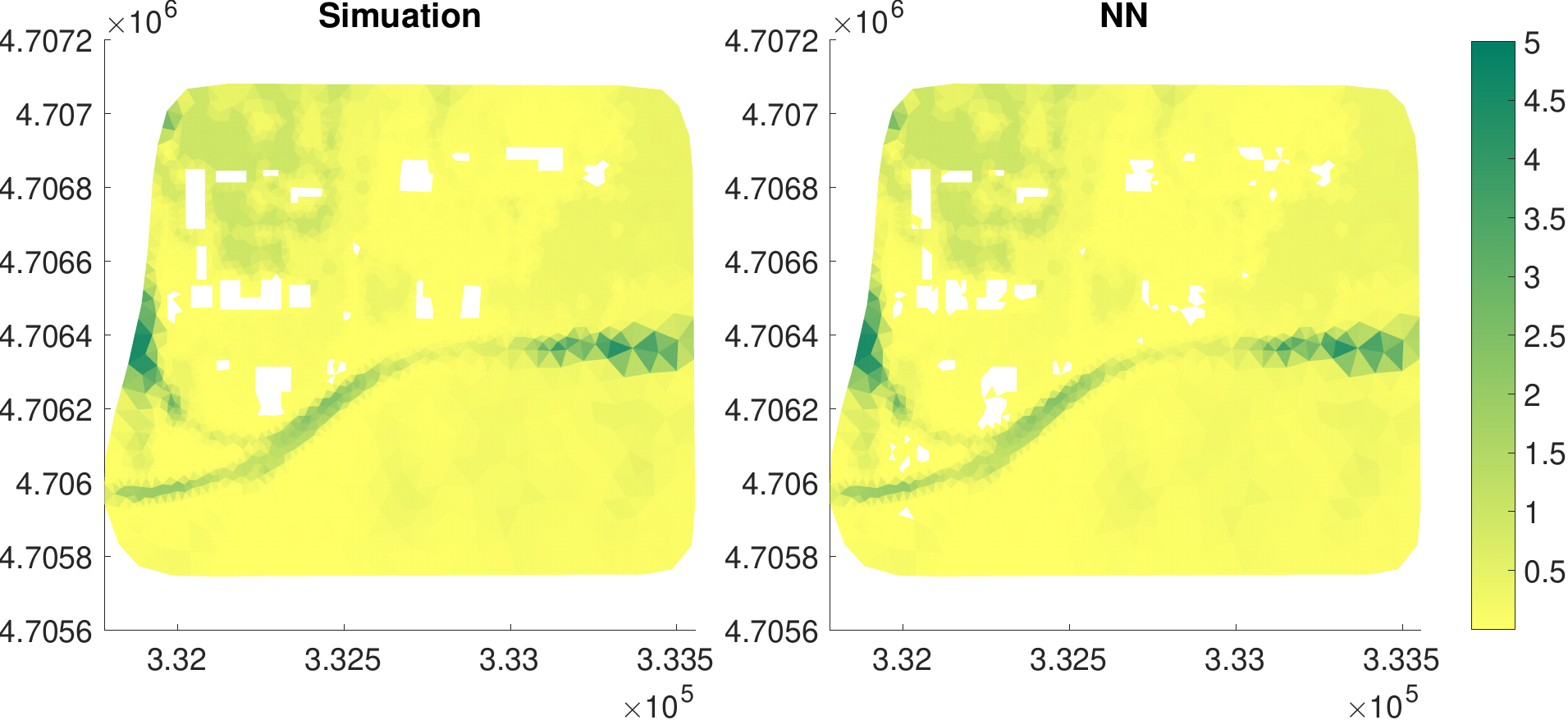}}%
    \subfloat[11 August 2014, 23:45]{%
        \includegraphics[width=0.47\textwidth]{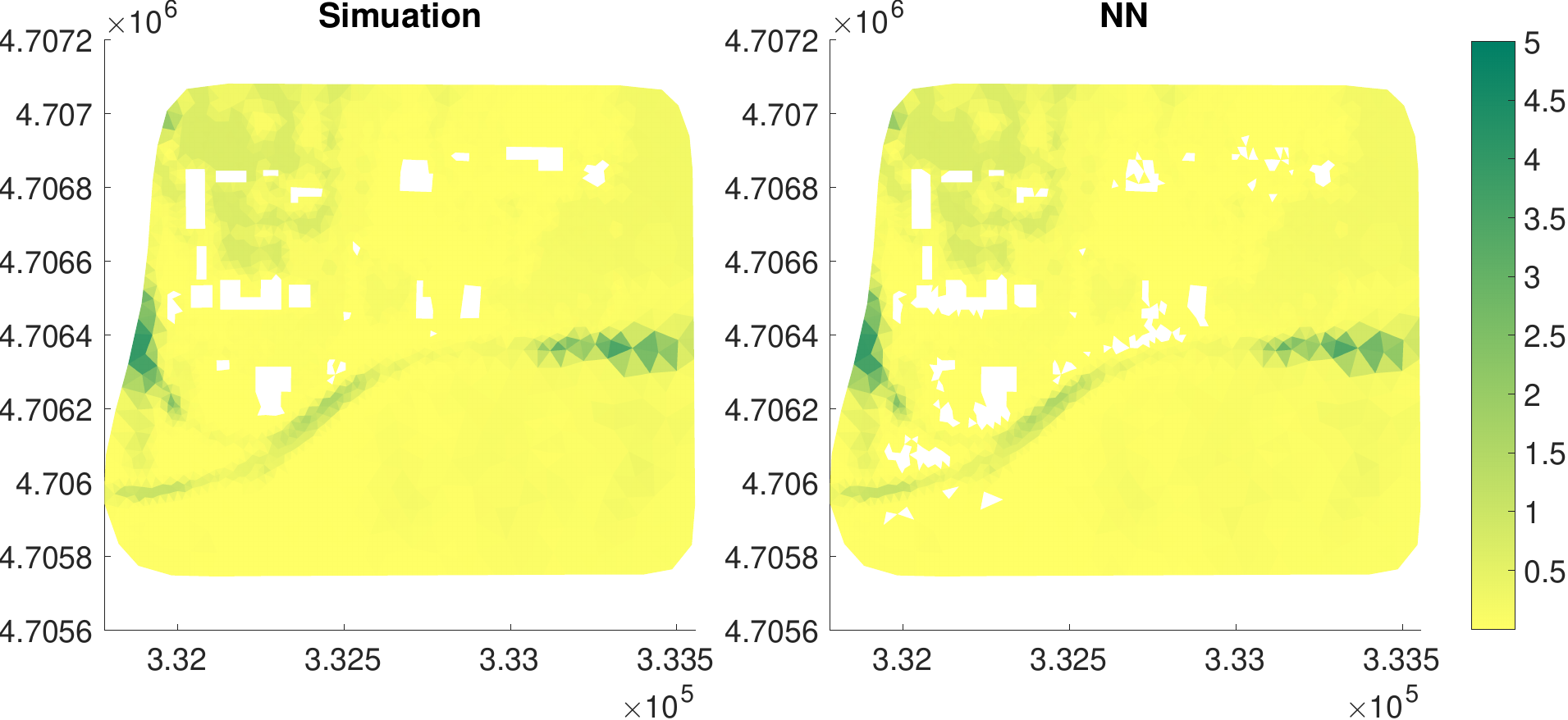}}
    \caption{Comparisons between tRIBS-Urban and $\widehat h$ at one parameter realization. Each panel places the simulator and surrogate fields side by side; water depths are in meters.}
    \label{fig:comparison_surrogate}
\end{figure}

\Cref{fig:comparison_surrogate} compares tRIBS-Urban and surrogate inundation fields for one parameter realization at two times. For the two displayed comparisons, the root-mean-square errors are of order $10^{-2}\,\mathrm{m}$.

\section{Sobol' variance decomposition and sampling}\label{app:sobol}

Let $W=\widehat h(x,t;\Param)$, $D=\operatorname{Var}(W)$, and $n_p=6$. Under the independent prior in \eqref{e:parameter_prior}, the Sobol' decomposition is
\begin{align}
    D
    =\sum_i D_i+\sum_i\sum_{j>i} D_{ij}+\cdots+D_{1\ldots n_p},
    \label{e:sobol_decomposition}
\end{align}
where
\begin{align}
    D_i
    &=\operatorname{Var}_{\Param_i}
      \!\left(\EE_{\Param_{-i}}[W|\Param_i]\right),\\
    D_{ij}
    &=\operatorname{Var}_{\Param_i,\Param_j}
      \!\left(\EE_{\Param_{-i,-j}}[
      W|\Param_i,\Param_j]\right)-D_i-D_j.
\end{align}
The first-order and total-order indices are
\begin{align}
    S_i
    &=\frac{D_i}{D},\\
    S_{T_i}
    &=\frac{\EE_{\Param_{-i}}
        \!\left[\operatorname{Var}_{\Param_i}(W|\Param_{-i})\right]}{D}.
    \label{e:sobol_total}
\end{align}
The total-order index sums the main effect and every interaction containing $\Param_i$, while the difference $S_{T_i}-S_i$ summarizes the normalized interaction contributions involving that parameter.

The estimates use Sobol' quasi-Monte Carlo sequences with base sample size $M=10{,}000$. For $n_p=6$, a first- and total-order Saltelli construction without second-order estimation requires $(n_p+2)M=80{,}000$ surrogate evaluations per analyzed output. 

\section{Expected-information estimators}\label{app:oed_computation}

Parameter EIG and prediction EIG are conditional mutual information quantities. This appendix gives the numerical procedures used to evaluate them across candidate locations.

\subsection{Nested Monte Carlo estimation of parameter EIG}
\label{app:NMC}

For each candidate location $\design$, draw outer samples $\param^{(i)}\sim p(\param)$ and
$y^{(i)}\sim p(y|\param^{(i)},\design)$. Draw independent inner samples $\param^{(i,j)}\sim p(\param)$. With outer and inner sample sizes $N_{\mathrm{out}}$ and $N_{\mathrm{in}}$, respectively, the nested Monte Carlo estimator is~\citep{Ryan2003}
\begin{align}
    \widehat U_{\Param}^{\mathrm{NMC}}(\design)
    =\frac{1}{N_{\mathrm{out}}}
    \sum_{i=1}^{N_{\mathrm{out}}}
    \Bigg\{
    \log p\!\left(y^{(i)}|\param^{(i)},\design\right) 
    -\log\!\left[
      \frac{1}{N_{\mathrm{in}}}
      \sum_{j=1}^{N_{\mathrm{in}}}
      p\!\left(y^{(i)}|\param^{(i,j)},\design\right)
      \right]
    \Bigg\}.
    \label{e:PO_NMC}
\end{align}
The inner average approximates the prior-predictive density $p(y^{(i)}|\design)$, allowing the likelihood of each simulated observation to be compared with its probability before the parameter realization is known.

\subsection{Variational estimation of prediction EIG}
\label{app:variational}

Let $q(z|y,\design;\lambda)$ approximate the posterior-predictive density. The Barber--Agakov lower bound in \eqref{e:GO_UL} can be written as
\begin{align}
    U_L(\design;\lambda)
    =\iint
      p(z|y,\design)
      p(y|\design)
      \log\!\left[
      \frac{q(z|y,\design;\lambda)}{p(z)}
      \right]
      \,\mathrm{d}z\,\mathrm{d}y.
    \label{e:GO_UL_integral}
\end{align}
The conditional density uses normalizing flows composed of invertible coupling transformations whose parameters are produced by neural networks~\citep{Dong2025}. Variational fitting chooses $\lambda$ to make $q(z|y,\design;\lambda)$ approximate the posterior-predictive density, while the fitting strategy determines how the density model shares information across candidate locations.

Draw $\param^{(i)}\sim p(\param)$ and
$y^{(i)}\sim p(y|\param^{(i)},\design)$, and set
$z^{(i)}=H(\param^{(i)})$. These samples give the Monte Carlo estimator
\begin{align}
    \widehat U_L(\design;\lambda)
    =\frac{1}{N}\sum_{i=1}^{N}
      \left[
      \log q\!\left(z^{(i)}|y^{(i)},\design;\lambda\right)
      -\log p\!\left(z^{(i)}\right)
      \right].
    \label{e:GO_UL_MC}
\end{align}
The notation $\widehat U_L(\design)$ in the main text suppresses the fitted value of $\lambda$ and any dependence of that value on $\design$. The calculation retains the prior-predictive term $-\log p(z)$, so $\widehat U_L$ represents the complete variational criterion rather than only a relative ranking score.

\section{Additional design diagnostics}\label{app:diagnostics}

The following figures provide qualitative diagnostics, with each comparison conditioning on one synthetic observation for its displayed location and target. They illustrate how a selected measurement updates the prediction, while additional synthetic observations would extend the assessment across possible data realizations.

\begin{figure}[!htb]
    \centering
    \includegraphics[width=\textwidth]{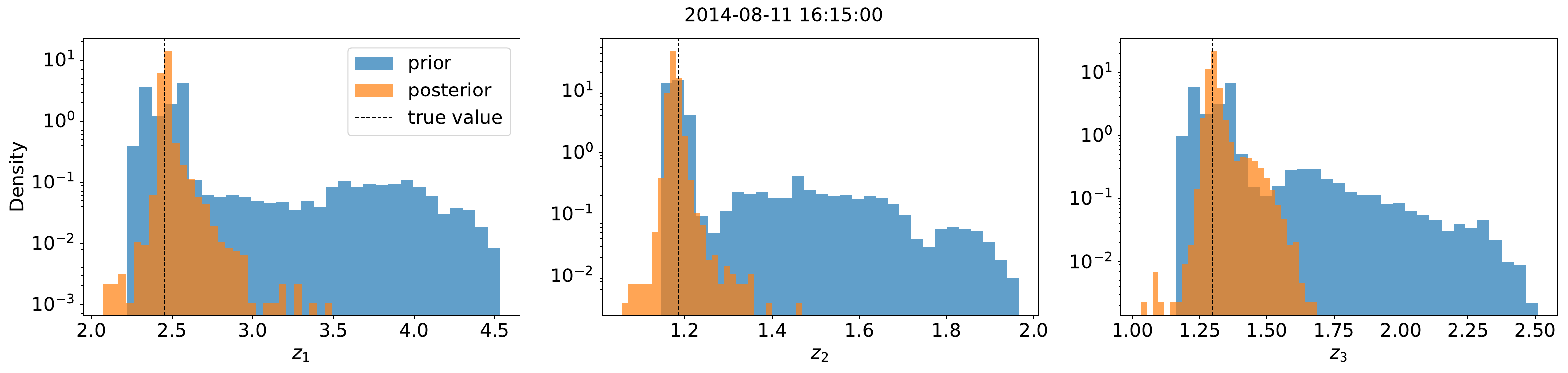}
    \caption{For one synthetic observation, the joint-target max-EIG location contracts the three displayed posterior-predictive distributions under the surrogate working model. Prior-predictive densities are blue, posterior-predictive densities are orange, density axes use logarithmic scales, and reference lines mark the simulated values.}
    \label{fig:GO_OED_posterior_multiple_locs_single_time}
\end{figure}

\begin{figure}[!htb]
    \centering
    \includegraphics[width=0.32\textwidth]{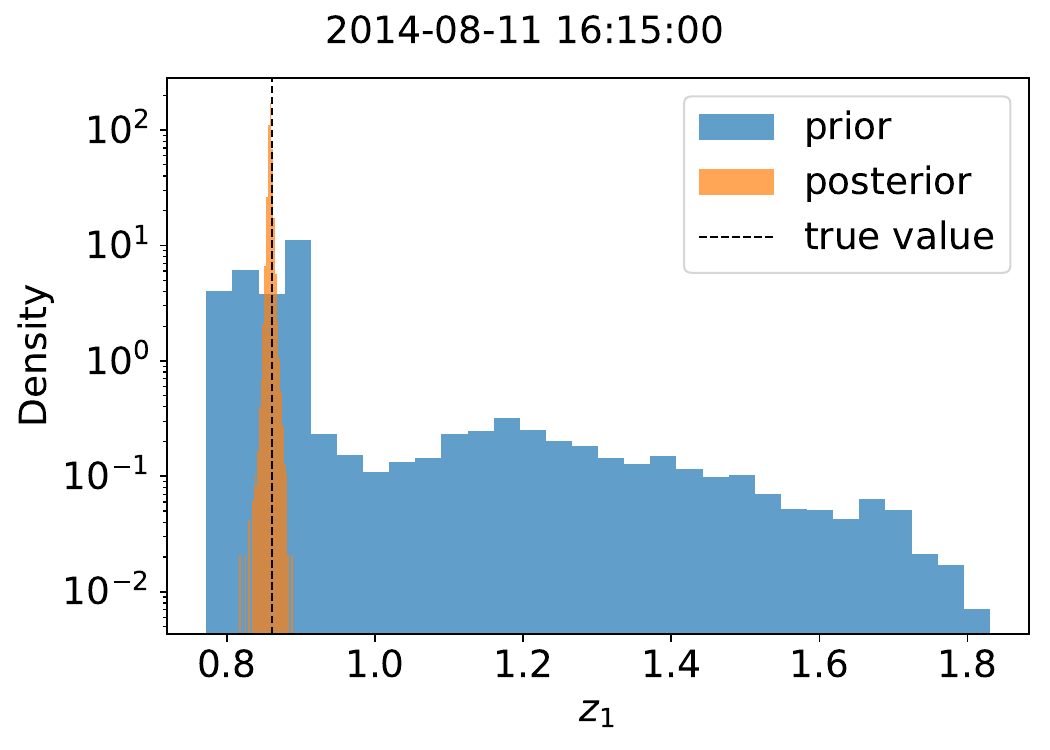}
    \includegraphics[width=0.32\textwidth]{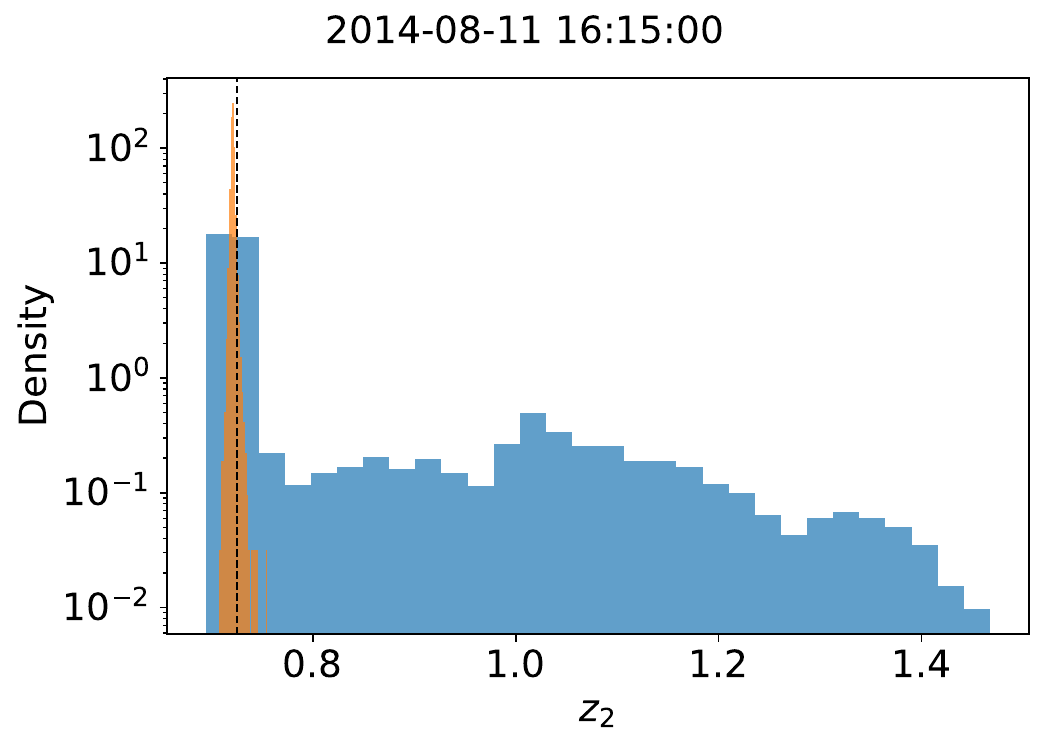}
    \includegraphics[width=0.32\textwidth]{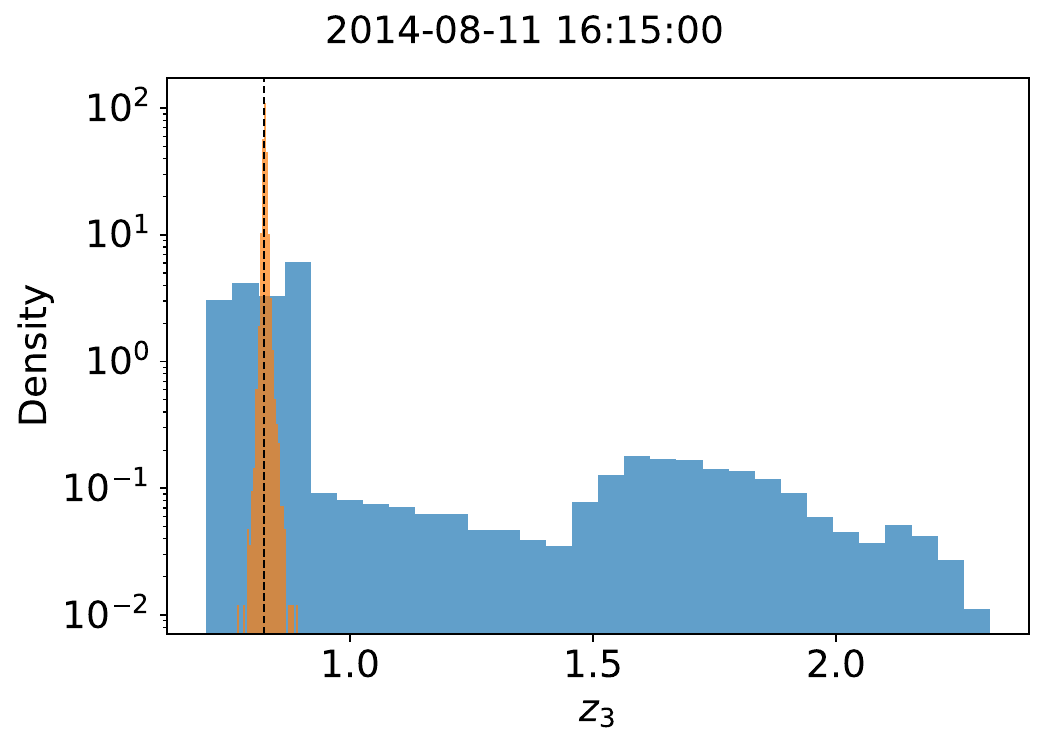}
    \caption{For one synthetic observation per target, the nonlocal max-EIG locations contract the displayed posterior-predictive distributions for regional averages in Regions 1--3 from left to right. Prior-predictive densities are blue, posterior-predictive densities are orange, density axes use logarithmic scales, and reference lines mark the simulated values.}
    \label{fig:GO_OED_posterior_spatial_avg}
\end{figure}